\documentclass[%
 reprint,
 amsmath,amssymb,
 aps,
]{revtex4-2}
\usepackage{float}
\usepackage{graphicx}
\usepackage{dcolumn}
\usepackage{bm}
\usepackage{hyperref}
\usepackage{physics}
\usepackage{makecell}
\usepackage{multirow}
\usepackage{array}
\usepackage{xcolor}
\usepackage{comment}

\begin{document}

\preprint{APS/123-QED}

\title{Surface Topological Quantum Criticality II:\\
Conformal manifolds, Isolated fixed points, and Entanglement}

\author{ Saran Vijayan}
\email{saran@phas.ubc.ca}
\author{Fei Zhou}
 
\affiliation{Department of Physics and Astronomy, University of British Columbia, 6224 Agricultural Road, Vancouver, BC, V6T 1Z1, Canada}
\date{\today}

\begin{abstract}
    In this article, we propose a possibility of realizing conformal manifolds, which are smooth manifolds formed by a family of scale-conformal invariant interacting Hamiltonians in two-dimensional quantum many-body systems. Such phenomena can occur in various interacting systems, including topological surfaces or 2D bulks. Based on a previous study, here we further demonstrate that a conformal manifold can emerge as an exact solution when the number of fermion colors, \(N_c\), in our models becomes infinite. We identify distinct exact marginal deformation operators uniquely associated with the conformal manifolds. When \(N_c\) is set to be finite but large, we also show that quantum fluctuations that induce fermion field renormalization can result in mildly infrared relevant or irrelevant renormalization-group (RG) flow within a conformal manifold. This leads to standard, isolated, infrared-stable Wilson-Fisher fixed points. These, along with ultraviolet stable fixed points, form a discrete manifold, due to the spontaneous symmetry breaking of an emergent \(SO(\mathcal{N})\) dynamical symmetry in the RG flow as \(N_c \rightarrow \infty\). In addition, we find that along the direction of the RG flow within the manifold, an EPR-like entanglement entropy in the fermion flavor space always increases. The infrared-stable Wilson-Fisher fixed points, induced by quantum fluctuations, are always linked to theories on the conformal manifold where interaction operators are maximally entangled. Our studies provide an effective framework for addressing topological quantum critical points with high-dimensional interaction parameter spaces that potentially host an overwhelmingly large number, often an exponentially large number of fixed points of complex stabilities. They also highlight the central role of entangled conformal operators and their entropy in shaping the universality classes of topological quantum phase transitions on surfaces or in bulk. We conclude our studies with open questions and possible future directions.
\end{abstract}

\maketitle

\section{Introduction}

One of the best-known features of a symmetry-protected state (SPT) \cite{ludwig(2008),kitaev(2009), gu(2009), ludwig(2010), gu(2012), wen(2013),zirnbauer(2016)} is the presence of distinct symmetry-protected gapless boundaries. These boundaries are robust and usually remain gapless as well-defined boundary phases when subject to weak symmetric perturbations, i.e., perturbations are invariant under the protecting symmetry group $G_p$.

Two well-known paths can lead to gapped boundaries. One is by conventional spontaneous symmetry breaking of protecting symmetries $G_p$ at boundaries (but not in the bulk)\cite{xu(2010),herbut(2013),sondhi(2013),Ponte(2014),Grover(2014),neupert(2015)}. The other approach involves developing topological order at boundaries without breaking the protecting symmetries\cite{fidkowski(2010),fidkowski(2011),senthil(2013),fidkowski(2013),metlitski(2015),senthil(2015)}. Whether local interactions can lead to the latter exotic scenario remains to be further researched. In this article, just as in a previous study of ours\cite{saran(2025)}, we will exclusively focus on the former path. We will study surface topological quantum critical points (sTQCPs) or phase boundaries that separate a gapless symmetry-protected boundary and gapped surface states where the protecting symmetry $G_p$ is broken spontaneously. Theories of these surface TQCPs define unique universality classes of the corresponding quantum
phase transitions if they do occur, and a thorough understanding of this issue can shed light on potential applications of topological states and their boundary physics in modern quantum technologies.

A fascinating application of SPT boundary states is in the context of emergent super-symmetries (SUSY) and supersymmetric conformal field theories. These had been explored in a few fascinating attempts to realize these
states in topological surfaces, where a single copy of SUSY theory appears on each surface of a 3D bulk\cite{Ponte(2014),Grover(2014),maciejko(2016),yao(2017),yao(2017)(2),yao(2018)}. An early effort in a two-dimensional lattice illustrates that two copies of kinematically decoupled SUSY theories can also coexist simultaneously\cite{SSLee(2007)}.

Recently, it was discovered that SUSY can also be present at a three-dimensional TQCP, where a single copy of SUSY conformal fields appears in the three-dimensional bulk as a result of the emergence of single Weyl fermion dynamics in the infrared limit\cite{zhou(2022),zhou(2023),zhou(2024)}. Currently, its intimate connection to lattice chiral fermions has also been the subject of thorough and intensive investigation. 

In addition, both the degrees of freedom and dynamic critical exponents at TQCPs can depend on interaction strengths and generally shall be treated as emergent infrared properties. In contrast to the more conventional paradigm of quantum criticality, these properties at TQCPs are not uniquely determined by protecting symmetry groups or topologies\cite{zhou(2021),zhou(2025)}.

It is generally believed that physics at TQCPs can be holographic, related to surfaces of higher-dimensional bulk with higher lattice symmetries\cite{dunglee(2019), zhou(2024)}. The emergent symmetries and T'Hooft anomalies strongly suggest such a paradigm. The surfaces of SPT typically carry 
gauge anomalies, playing critical roles in our current studies of topological matter and TQCPs\cite{wen(2013)}.

Beyond its application to SUSY, it was recently pointed out that gapless surface fermions can serve as an ideal platform to explore quantum dynamics related to conformal manifolds, a smooth manifold of scale-invariant and conformal interactions \cite{saran(2025)}. Traditionally, this subject has been extensively studied in the context of AdS/CFT, where conformal field theories in $3+1$ dimensions are upgraded to superconformal ones that naturally support exact marginal deformation operators \cite{leigh(1995),maldacena(1999),ooguri(2024)}. The earlier context in which conformal manifolds were introduced and studied is a family of $ 1+1$-dimensional conformal field theories \cite{zamolodchikov(1986),zamolodchikov(1987)}. A family of Luttinger liquids can also be thought to form a non-compact manifold, with their properties recently quite well understood\cite{wen(2020)}. 

As far as the authors are aware of, the recent discussions on the potential existence of conformal manifolds in $ 2+1$D topological surfaces are one of the few rare efforts to search for such smooth manifolds in scale-conformal symmetric theories, but with several unanswered open questions. It is for this reason that we have decided to examine this issue with a more advanced technique, hoping to offer better insight into such a possibility. 

Most of our discussions below are not restricted to topological surfaces but are applicable to other $2 + 1D$ bulk gapless fermions with relativistic space-time symmetries. On the other hand, topological surfaces may be one of the few most practical quantum condensed matter systems to realize such fascinating phenomena, given that their stability comes with the protecting symmetries. But again, the phenomena discussed below are not limited to topological surfaces. 

Our main objective here is to illustrate that conformal manifolds can be a very powerful approach to the study of sTQCPs in high spatial dimensions, such as two-spatial-dimensional quantum systems that are neither exactly solvable nor integrable, nor supported by super-conformal algebras, which often naturally imply the existence of exactly marginal deformation operators. 

This is especially true when we aim to understand sTQCPs in the multi-dimensional interaction parameter space, with its dimension $D_p$ (where the subscript $ p$ refers to interaction parameters). Given that they can, in principle, appear in high spatial dimensions without the usual super-conformal algebras or integrability, it is appealing to pinpoint their existence in interacting topological states, even though their existence does require taking an extreme limit of our models, as we will see below. 

Furthermore, the basic concept of conformal manifolds also provides an extremely valuable starting point for practical studies of more realistic limits of sTQCPs. Along this path, we can then further connect what happens within the conformal manifold in the presence of controllable quantum fluctuations with another fascinating aspect, an increase in entanglement under the infrared flow of the renormalization-group equations (RGEs). 

In a recent previous research of topological surface quantum criticality, the authors have demonstrated that topological surface quantum critical points (sTQCPs) that separate weakly interacting gapless surfaces of a fermionic symmetry protected state (SPT) from a gapped surfaces that spontaneously symmetry break a protecting symmetry $G_p$ can be closely related to a conformal manifold, i.e. a smooth manifold of scale-invariant fixed points. Such manifolds support a unique number of exact marginal deformation operators when higher-order quantum fluctuations are neglected and manifolds are stable. 

Within the interacting models in that previous study, the phase boundary (see also below) is a codimension-one sub-manifold embedded in the interaction parameter space of $D_p$ dimensions. The dimension of the phase boundary manifold is $D_p - 1$. For $D_p = 3, 6$, the phase boundaries are 2-dimensional cone-like surfaces and 5-dimensional sub-manifolds, respectively. 

However, not all the interactions on the phase boundaries are exactly scale-invariant, and most interacting models on the phase boundaries further flow, in the limit of infrared, into sub-manifolds under the scale transformation or under the influence of the renormalization group equations (RGEs). 

It turns out that conformal manifolds naturally emerge at a one-loop level of renormalization group equations as a sub-manifold embedded in the co-dimension one phase boundaries. For instance, for $D_p = 3$ and a 2-dimensional phase boundary, the conformal manifold was obtained as a one-dimensional ring, isomorphic to $S^1$. In contrast, for $D_p = 6$ and a 5-dimensional phase boundary, the conformal manifold was derived to be isomorphic to a two-sphere surface, $S^2$. 

However, higher-order quantum fluctuations can further induce flows within these manifolds, which break the manifolds down to more conventional, well-isolated Wilson-Fisher fixed points. At the most stable infrared fixed points, it was observed that the interaction operators always appear to have maximally entangled structures. In our studies of attractive pairing interactions, this entangled structure provides further evidence of an explicit connection to Einstein-Podolsky-Rosen bipartite entanglement. 

These puzzling structures, on the one hand, are extremely fascinating; on the other hand, they demand more careful investigation. This has inspired us to examine the interacting Hamiltonians appearing on those manifolds more thoroughly, with finer resolutions and a better characterization of the different interacting models belonging to those smooth manifolds. 

To achieve these objectives, we first need to upgrade our approaches to interacting Hamiltonians in our previous studies. The 4-fermion interacting model employed in our earlier research has an upper critical dimension at $(1 + 1)D$. It usually can be considered to be an infrared (IR) effective field theory of an ultraviolet (UV) completed model with Yukawa coupling with an upper critical dimension equal to $(3 + 1)D$.

In the Yukawa model approach, emergent bosons, which are physically detectable, appear explicitly as fundamental fields along with the elementary fermionic fields, offering finer resolutions of the physical degrees of freedom near surface topological quantum criticality. In general, the Yukawa model approach not only reproduces the 4-fermion interaction picture below a bosonic mass gap $M_B$ but furthermore provides many detailed structures at the scale above the mass gap $M_B$. 

As $M_B$ needs to be set to zero at a surface topological quantum critical point (sTQCP), the energy window where the 4-fermion interacting model is applicable in principle becomes diminishing. This makes the 4-fermion model an inconvenient approach for detailed studies of sTQCPs. At the same time, this also indicates that the Yukawa model is a uniquely powerful approach for the full understanding of sTQCPs from UV to IR.

This is the focus of this article. We will reveal a few exciting new aspects at sTQCPs which are otherwise much more challenging to capture in the 4-fermion model, if not impossible. In the Yukawa coupling approach, both bosonic and fermionic fields appear explicitly as fundamental fields that further couple with each other. When bosons are massive with a well-defined mass gap $M_B$ in the IR limit, one can treat bosons as mediators inducing 4-fermion and other interactions. However, at an sTQCP where the $M_B$ is set to zero, both bosons and fermions are massless, leading to a conformal field theory which we will study in detail.

The main questions we are going to answer are:

\paragraph*{\textbf{Q1}:} Can sTQCPs that involve multiple-dimensional interaction parameter space, be characterized by a conformal manifold, i.e., a smooth manifold of scale-invariant interacting theories, with its dimension fixed by the number of \textit{exact marginal} deformation operators, rather than by simple isolated Wilson-Fisher fixed points?

\paragraph*{\textbf{Q2}:} Does a conformal manifold support only one universality class of quantum critical phenomena, or can it support multiple classes? If the whole manifold only supports one universality class, which we can call \textit{featureless} from the standard point of view of quantum criticality, what are the other structures or features that one can apply to differentiate different points on the manifold?

\paragraph*{\textbf{Q3}:} Under which conditions, such a conformal manifold, if exists, is infrared stable, and can it further break down to more conventional isolated fixed points under certain conditions?

\paragraph*{\textbf{Q4}:} What are the fundamentals behind the breakdown and RGE flow pattern within a conformal manifold? More specifically, what dictates the direction of infrared flow on the manifold if it happens to break down, say, due to some quantum fluctuations?

\section{Summary of our main results}

We summarize our results based on the analysis presented in the following sections. To better control our analysis and separate different types of quantum fluctuations at sTQCPs, we introduce a Yukawa coupling model with $\mathcal{N}$ flavors of 2-dimensional \textit{relativistic} gapless Dirac fermions with charge $e$, $\psi^{\dagger}_i$, $i=1,..,\mathcal{N}$. 

These flavors can model $\mathcal{N}$ Dirac cones on the surface of a cubic topological insulator (TI)\cite{fukane(2005),bernevigzhang(2006),bernevighugheszhang(2006),fukanemele(2007),moorebalents(2007),hasankanereview(2010), Qizhangreview(2011),bernevigbook(2013)} as a practical application. However, the main results of our studies are not restricted to this particular application. They are also valid for discussions of strongly interacting gapless fermions at general two-dimensional TQCPs. 

Each fermion flavor further couples to a complex bosonic field with charge $2e$, $\phi^{\dagger}$, through a Yukawa coupling with strength $g_{i}$,$i=1,..,\mathcal{N}$.
As stated above, all the fermionic fields are gapless, so that they can be surface fermions of an SPT state or fermions at bulk TQCPs. 
However, the bare mass of the complex boson field is introduced as a tunable parameter $M_B$(subscript $B$ refers to bosons). Only at tQCPs (or sTQCPs), the bare mass $M_B$ needs to be tuned to be a critical value (see below).

The parameter space of our theory is spanned by the boson mass $M_B$, and Yukawa coupling $\{ g_{i} \}$, $i=1,..,\mathcal{N}$. The dimension of the parameter space, therefore, is $D_p = \mathcal{N}+1$. 

Finally, we also assume each flavor of fermions has $N_c$ colors, and we will study the model in the limit of large $N_c$. Note that for a given $i$th flavor, fermions with different colors interact with the bosonic $\phi$-field via the same coupling $g_i$, i.e, $g_i$ only depends on flavor indices $i=1,..,\mathcal{N}$ but is independent of colors. We further perform our calculations near the upper critical dimensions $(3+1)D$ at $d = 3 - \epsilon$, with $\epsilon \ll 1$, to achieve rapid convergence. 

\paragraph*{\textbf{A1}:} We show that in the limit of infinite $N_c$, sTQCPs in our model are precisely characterized by a conformal manifold. For $\mathcal{N}$ flavors, the manifold is given by,
\begin{eqnarray}
    \sum^{\mathcal{N}}_{i=1} g^2_i = \epsilon, \,\, M^2_{B} = \sum^{\mathcal{N}}_{i=1} g^2_{i} = \epsilon,\label{smmrylrgeNc}
\end{eqnarray}
which is a sphere embedded in $\mathcal{N}$-dimensional flavor space, that is isomorphic to $S^{\mathcal{N}-1}$.

Eq. (\ref{smmrylrgeNc}) shows that the mass $M_{B}$ has to be fine-tuned to be critical. All interacting Hamiltonians are forced to be on a ($\mathcal{N}-1$)-dimensional sphere with radius given by $\sqrt{\epsilon}$. 

This is a conformal manifold with co-dimension equal to two, i.e. $D_{co}=2$ (the subscript \textit{co} refers to the co-dimension), embedded in the parameter space with the dimension $D_p=\mathcal{N}+1$.

As a comparison, we also show the mean-field phase boundary of the model that is given as (without self-interactions),
\begin{eqnarray}
    M^2_{B} = \sum^{\mathcal{N}}_{i=1} g^2_{i} \label{smmryMFT}.
\end{eqnarray}

 Here $M_B$ is not constrained to a specific value, in contrast with Eq. (\ref{smmrylrgeNc}). The phase boundary, as expected, is simply a codimension one manifold embedded in a $D_p=\mathcal{N}+1$-dimensional parameter space spanned by a) Yukawa coupling parameters $\{g_i\}$, $i=1,..,\mathcal{N}$, and b) the mass operator $M_B$. 
 In the parameter space of $D_p=\mathcal{N}+1$ dimensions, this phase boundary manifold has a codimension one, one dimension lower than the codimension of the conformal manifold.

Comparing Eqs. (\ref{smmrylrgeNc},\ref{smmryMFT}), We again find that the conformal manifold isomorphic to an $S^{\mathcal{N}-1}$ is a sub-manifold in the $\mathcal{N}$-dimensional phase boundary manifold, where the radius is precisely fixed at a value given by the spatial dimension $3-\epsilon$. This is a general conclusion of our Yukawa coupling models, independent of the number of flavors $\mathcal{N}$.

\paragraph*{\textbf{A2}:} The whole sphere-like conformal manifold isomorphic to $S^{\mathcal{N}-1}$ only supports one single set of scaling dimensions or conformal dimensions. That is, all operators, relevant, irrelevant, and marginal at different points of the manifolds, share the same scaling/conformal dimensions, respectively. From the traditional quantum criticality point of view, the manifold is featureless and has much less structure compared with K$\Ddot{a}$hler manifolds in the superconformal Yang-Mills theory\cite{seiberg(1988),seiberg(1994)}. This aspect was also pointed out in our previous
studies.

However, the microstructures of these operators differ significantly, as was also noted in a previous study. In this article, we further quantify the difference by applying a measure of entanglement. The differences can be visualized most explicitly by introducing a flavor-space entanglement entropy, which effectively captures the emergence of EPR-like entanglement in the Yukawa interaction operators of strength, $\{g_i \}$, $i=1,..,\mathcal{N}$.

\paragraph*{\textbf{A3}:} At a finite $N_c$, the conformal manifold breaks down to isolated Wilson-Fisher fixed points with mild or slow RGE flow within the manifold. The dominating quantum fluctuations that drive such RGE flow are due
to the fermion field renormalization effects, which typically explicitly break the emergent $SO(\mathcal{N})$ symmetry present in the limit of $N_c = +\infty$.

For $\mathcal{N} = 2$, the manifold breaks down to a total $3^2 -1$ or $8$ fixed points, four of which are infrared stable ones
forming a fundamental representation of $Z_2\otimes Z_2$. The other four are infrared unstable ones which emit flows toward the stable ones (see table \ref{ykwaN2tble} or Fig. \ref{RGflowlrgeNcN2}). For $\mathcal{N}=3$, the manifold breaks down
to a total of $3^3 - 1$ or 26 fixed points. Eight of them
are infrared stable ones, while the remaining 18 are infrared
unstable ones, each of which has a certain number of
relevant operators associated with it (see table \ref{ykwaN3tble}). All the stable
fixed points have the highest symmetric form of,

\begin{eqnarray}
    g^2_{i,c} = \frac{\epsilon}{\mathcal{N}} + O(\epsilon^2,1/N^2_c),\,\,\, i = 1,.., \mathcal{N}.
\end{eqnarray}
which indicates that $\prod_{i=1,.,\mathcal{N}} g^2_i \neq 0$.

\paragraph*{\textbf{A4}:} It appears that in all our studies, the RGE flow in the limit of finite but large $N_c$ always directs toward the fixed points where the interaction operators located at a conformal manifold have maximally entangled structures (see Fig. \ref{RGflowlrgeNcN2}). In this article, we will
mainly focus on the limit of $\mathcal{N}=2,3$, overlapping with
our previous 4-fermion model studies.

\paragraph*{\textbf{A5}:}
As a by-product of our studies, we also comment that a SUSY fixed point or conformal field theory can also appear in our conformal manifold. However, in all our studies with multiple flavors of Dirac fermions or Dirac cones with $\mathcal{N} = 2,3$, they are most unstable in the infrared limit once the subleading quantum fluctuations of order $O(1/{N_c})$ are included. In fact, the interaction operators at those fixed points are least entangled with a minimum entanglement entropy of zero. Therefore, they don't play a significant role in our discussions of sTQCPs and their universalities in this article.

These conclusions above in \textbf{A1-A5} here are generic. The result for an arbitrary $\mathcal{N}$, especially for a large $\mathcal{N}$ limit, will be presented in a separate article to appear soon.

\section{Recap: \\
Conformal manifolds and phase boundary of $\mathcal{N}$-flavor TI surface studied using 4-fermion theory}

Below, we briefly review the results of the four-fermion interacting model studied in Ref.\cite{saran(2025)}. We summarize the main conclusions in subsections 1-4. For detailed discussions, we refer to the original article \cite{saran(2025)}.

\subsubsection{Effective field theory}  In \cite{saran(2025)}, we showed that the surface of a cubic crystal of a 3d TI can host multiple flavors of 2-component Dirac fermions. The surface Brillouin zone of such a cubic crystal has four distinct time-reversal invariant points that can host these multiple Dirac cones. The number of surface fermion flavors is stable as long as the crystal symmetries are respected.  

In this context, we consider a 3d TI with crystal symmetry in addition to its protecting symmetries, ensuring that the multiple surface Dirac cones remain stable. Let $\mathcal{N}$ represent the number of surface Dirac fermion flavors for a given bulk. We introduce a short-ranged attractive interaction to the  $\mathcal{N}$-flavor TI surface. The effective low-energy theory is then given by \cite{saran(2025)},
\begin{eqnarray} 
    \mathcal{S}_{4f} &=& \int d^{d}\textbf{r}d\tau  \,\, \mathcal{L}_{4f} \label{Lfrfmn} \\ 
    \mathcal{L}_{4f} &=& \sum^{\mathcal{N}}_{i=1} \psi^{\dagger}_{i}\left[\partial_{\tau} + i s_y \partial_x - i s_x \partial_y  \right] \psi_i \nonumber\\
     &-& \sum^{\mathcal{N}}_{i,j=1} V_{ij} \psi^{\dagger}_{i} \left(-i s_y \right)\psi^{\dagger T}_{i} \psi^{T}_{j}\left(i s_y \right) \psi_{j}\nonumber 
\end{eqnarray}

Here $s_{i}(i=x,y,z)$ are Pauli matrices defined in spin-1/2 space. $\psi^{\dagger}_{i}$ is a 2d 2-component fermion field operator of flavor '$i$' with the definition $\psi^{\dagger}_{i} = (\psi^{\dagger}_{i,\uparrow} \,\,\, \psi^{\dagger}_{i,\downarrow})$. $V_{ij}$ is the matrix element of an $\mathcal{N}\times \mathcal{N}$ symmetric interaction matrix, $\hat{V}$. Thus, the parameter space of theory is a $D_p= \mathcal{N}(\mathcal{N}+1)/2$-dimensional space.
Physically, it determines the strength of scattering of a singlet fermion pair from a surface Dirac cone of flavor '$i$' to flavor '$j$'. 

\subsubsection{Mean-field analysis: Phase boundary manifold} We use mean-field theory to examine the phase boundary of the gapless phase of the $\mathcal{N}$-flavor TI surface in the $D_p$-dimensional parameter space. Let us define the order parameter as follows:
\begin{eqnarray}
    \triangle_{i} = \sum^{\mathcal{N}}_{j=1} V_{ij} \sum_{\textbf{k}}\left<\psi^T_{j,\textbf{k}} \left( is_y \right) \psi_{j,-\textbf{k}}\right>
\end{eqnarray}
We essentially have an $\mathcal{N}$-component complex order parameter field for the $\mathcal{N}$-flavor TI surface. 

Let $\hat{V}_{p}$ be the interaction matrix that describes the phase boundary where the solutions to the order parameter equation above approach zero. Then $\hat{V}_p$ must obey,
\begin{eqnarray}
     \det \left[\hat{V}_p - \frac{\Lambda}{2\pi} \hat{I}_{\mathcal{N}} \right] = 0 \label{phsebndrmnfldgen}
\end{eqnarray}
where $\Lambda$ is the UV cut-off of the interaction. and $\hat{I}_{\mathcal{N}}$ is an identity matrix of size $\mathcal{N}$. The solution to this equation defines the phase boundary. Since we have one equation and $D_{p}$ unknowns, the solution will describe a ($D_{p}-1$)-dimensional manifold. In conclusion, the phase boundary forms a ($D_{p}-1$)-dimensional manifold.

\begin{figure}[b]
\includegraphics[width=8.5cm, height=8.1cm]{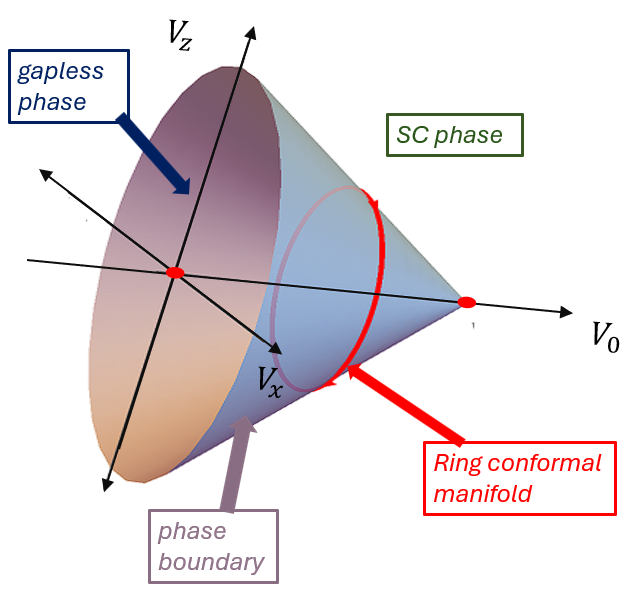}
\caption{Phase boundary (Eq. (\ref{phsebndrmnfldN2})) separating the gapless phase from the gapped superconducting phase for the $\mathcal{N}=2$ TI surface in the parameter space spanned by $(V_0, V_x, V_z)$ (Eq. (\ref{Vdef})). The red ring denotes the conformal manifold (Eq. (\ref{4fN2fp})) embedded on the phase boundary dictating its universality class. Red points denote isolated fixed points. Image from Ref.\cite{saran(2025)}}
\label{phsebndry_4frmn}
\end{figure}

Let us understand the geometry of the phase boundary for the simplest case of $\mathcal{N}=2$ flavor TI surface. Here, the parameter space is $3$-dimensional. For convenience, we reparametrize the interaction matrix as 
\begin{eqnarray}
  \hat{V} = V_0 \hat{I}_{2} + V_x \sigma_x + V_z \sigma_z  \label{Vdef}
\end{eqnarray}
 where $\sigma_i(i=x,z)$'s are Pauli matrices. Plugging this back into Eq. (\ref{phsebndrmnfldgen}), we get the following equation for the phase boundary,
\begin{eqnarray}
    V^2_{z,p} + V^2_{x,p} = \left(\frac{\Lambda}{2\pi} - V_{0,p} \right)^2. \label{phsebndrmnfldN2}
\end{eqnarray}
It describes the shape of a cone with tip at $V_0 = \Lambda/2\pi$ in the parameter space spanned by $(V_0, V_x, V_z)$ , as shown in Fig. \ref{phsebndry_4frmn}. The system remains gapless if the interaction parameters lie inside the cone, while it is in a superconducting phase outside the cone.

\subsubsection{Renormalization group analysis: Conformal manifolds} In \cite{saran(2025)}, we performed renormalization group analysis of the theory in Eq. (\ref{Lfrfmn}). We summarize the results here. In $d = 1 + \epsilon$ dimensions and at one-loop level, the RGE is,
\begin{eqnarray}
    \frac{d\hat{V}}{dl} &=& - \epsilon \hat{V} +  \hat{V}^2 + O(V^3)
\end{eqnarray}
where we rescaled $\dfrac{8\hat{V}}{\pi}\rightarrow \hat{V}$ for convenience. By increasing $l$, we move towards the infrared (IR) limit. The RGE is invariant under the $SO(\mathcal{N})$ rotations of $\hat{V}$, while the action in Eq. (\ref{Lfrfmn}) is not. Consequently, in addition to the isolated fixed points, we find conformal manifolds as solutions of the beta function above. 

The key finding was that these conformal manifolds form lower-dimensional sub-manifolds embedded on the bigger phase boundary manifold in Eq. (\ref{phsebndrmnfldgen}). Below, we shall demonstrate this using the example of $\mathcal{N}=2$ TI surface.

For $\mathcal{N}=2$, the conformal manifold is given by,
\begin{eqnarray}
    V^2_{z,c} + V^2_{x,c} = \frac{\epsilon^2}{4},\,\, V_{0,c} = \frac{\epsilon}{2}.    
    \label{4fN2fp}
\end{eqnarray}
which takes the shape of a ring in the parameter space. The subspace spanned by the marginal and the irrelevant couplings of $\hat{V}_c$ forms the phase boundary of the gapless phase. The geometry of the phase boundary is found to be exactly conical in shape (see Fig. (\ref{phsebndry_4frmn})), in agreement with the mean-field result in Eq. (\ref{phsebndrmnfldN2}). 

However, the renormalization group (RG) analysis provided further insights into the infrared physics at the phase boundary. RGE tells us that the ring conformal manifold in Eq. \ref{4fN2fp} embedded on the phase boundary dictates its universality class, except for its tip, which is an unstable isolated fixed point.

To summarize, we used mean-field analysis to identify the phase boundary manifold (as shown in Eq. \ref{phsebndrmnfldgen}) of the interacting $\mathcal{N}$-flavor TI surface. Meanwhile, the one-loop renormalization group analysis demonstrated that the lower-dimensional conformal submanifolds (described in Eq. \ref{4fN2fp} for $\mathcal{N}=2$) determine the universality classes of states in these larger phase boundary manifolds.


\subsubsection{Renormalization group analysis beyond leading order: breakdown of a conformal manifold} 
 The two-loop fermion field renormalization effect breaks the $SO(\mathcal{N})$ symmetry of the RGE. This generates an RG flow along the conformal manifold, causing the manifold to break into isolated fixed points.
 
 To study the RG flow along the ring manifold for $\mathcal{N}=2$, we define, $$\tan \phi = V_z/V_x.$$ Then,
 \begin{eqnarray}
    \frac{d\phi}{dl} = -2a V^2_0 \sin 2\phi
\end{eqnarray}
where $a>0$ is a numerical constant. The IR unstable fixed points on the phase boundary are at $\phi = \pi/2,3\pi/2$, and the IR stable fixed points are $\phi = 0, \pi$. 

Following Eqs. (\ref{Vdef},\ref{4fN2fp}), at the IR unstable fixed points, the fermions only have intra-cone interactions, but without inter-cone interactions, as the $\hat{V}$ matrix is diagonal. 

However, at the IR stable points, in addition to the intra-cone scattering, there is strong inter-cone scattering because of the presence of off-diagonal components in the interaction matrix $\hat{V}$. This implies an emergence of entangled fields at the IR stable fixed points, which we will further explore below in this article.

\section{\label{sec:Ykwathry} Effective Yukawa Field Theory}

We observed that the conformal submanifolds determine the universality classes of the phase boundary manifold (Eq. (\ref{phsebndrmnfldgen})) when studied at the one-loop level. Among these, the infrared(IR) stable conformal submanifold on the phase boundary exhibits an intriguing feature: The critical interaction matrix $\hat{V}_c$ lying on this submanifold is factorizable. For instance, in the $\mathcal{N}=2$ case, one can infer from the manifold equation in Eq. (\ref{4fN2fp}) that the critical interaction matrix $\hat{V}_c$ on the submanifold can be recast in the following form:

\begin{eqnarray}
    \hat{V}_c = \epsilon\, \textbf{n}_2 \otimes \textbf{n}_2.
\end{eqnarray}
where $\textbf{n}_2$ is a 2-component unit vector and $\epsilon = d-1$. Here, $\hat{V}_c$ is a rank-2 tensor, and the symbol ‘$\otimes$’ denotes the tensor (outer) product of two vectors. A similar structure holds for the $\mathcal{N}=3$ case as discussed in Ref.\cite{saran(2025)}.

Consequently, this IR stable conformal submanifold can be effectively studied by an emergent single-boson Yukawa field theory, wherein a single flavor of complex boson interacts with $\mathcal{N}$-flavors of fermions. 
Since the Yukawa theory takes into account the quantum fluctuations of the massless bosons, it forms a better description of the surface topological quantum criticality on the strongly interacting $\mathcal{N}$-flavor TI surface. 

In this article, we conduct a renormalization group analysis of the emergent Yukawa field theory. To facilitate this, we introduce an internal color space of dimension $N_c$ for each fermion flavor and perform a double expansion in powers of both $\epsilon = 3-d$ and $1/N_c$. In this enlarged color-flavor space, we define the surface fermion spinor field $\psi_{i,\alpha}$,  where $i=1,..,\mathcal{N}$ labels the flavor and $\alpha = 1,..N_c$ labels the color. Note that $\psi_{i,\alpha}$ itself is a 2-component spinor in the spin-1/2 space.
The field theory now becomes,

\begin{eqnarray}
\mathcal{S}_{Y,N_c}[\phi,\psi_i] &=& \int d^{d}\textbf{r}d\tau \left[\mathcal{L}_{f,N_c} + \mathcal{L}_{b} + \mathcal{L}_{fb,N_c} \right] \label{YkwEFTNc} \\
    \mathcal{L}_{f,N_c} &=& \sum_{i=1}^{\mathcal{N}}\sum_{\alpha=1}^{N_c}\psi^{\dagger}_{i,\alpha} \left[ \partial_\tau + i s_y \partial_x - i s_x\partial_y \right]\psi_{i,\alpha}\nonumber \\ 
    \mathcal{L}_{fb, N_c} &=& \sum^{\mathcal{N}}_{i = 1} \sum_{\alpha=1}^{N_c}  \frac{g_i}{\sqrt{N_c}} \phi^{*}\,\psi^{T}_{i,\alpha} i s_y \psi_{i,\alpha} +\text{h.c.} \nonumber\\
    \mathcal{L}_b &=& |\partial_{\tau} \phi |^2 + \sum_{i=1}^{d}|\partial_{i} \phi |^2 + \lambda |\phi|^4 \nonumber
\end{eqnarray}

Here $\phi(\textbf{r},\tau)$ represents the massless complex boson field. $g_i$ gives the coupling strength between the boson and the $i$th spin singlet fermion pair, and $\lambda$ is the four-boson interaction strength.

  In the non-interacting limit, the action is invariant under the symmetry group $SU(\mathcal{N}) \otimes SU(N_c)$. However, the interaction term only possesses $SO(N_c)$ invariance. This limitation arises because the flavor-dependent coupling $g_i$ explicitly breaks the  $SO(\mathcal{N})$ symmetry. As a result, the interacting theory is only invariant under $SO(N_c)$.

 We begin by examining the renormalization group equation (RGE) in the limit as $N_c$ approaches infinity. Notably, the continuum of strong-coupling fixed points forms a conformal manifold isomorphic to $S^{\mathcal{N}-1}$ within an $\mathcal{N}$-dimensional parameter space. We find that the entanglement entropy of the operators on the conformal manifold depends on the manifold's coordinates. Next, we examine the finite $N_c$ effects that break this manifold at order $O(1/N_c)$. We demonstrate that the quantum fluctuations at the $O(1/N_c)$ level result in a renormalization group flow towards fixed points whose Yukawa interaction operators are maximally entangled.

\begin{figure}[b]
\includegraphics[width=7.0cm, height=9.8cm]{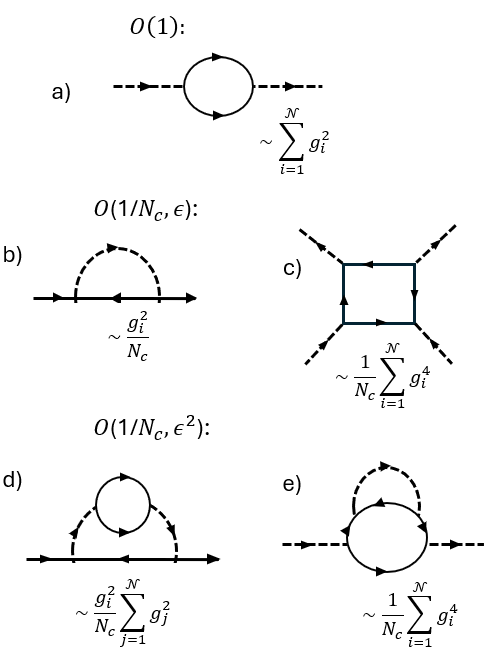}
\caption{Feynman diagrams that contribute to the RGE in Eq. (\ref{YkwaRGENcfntegen1},\ref{YkwaRGENcfntegen2}). a) represents the leading order diagram in the large $N_c$-expansion that remains finite in the limit of $N_c \rightarrow \infty$. b) and c) are subleading diagrams which is of the first order in the $\dfrac{1}{N_c}$-expansion, while d) and e) represent two-loop diagrams that are also of the first order in the $\dfrac{1}{N_c}$-expansion and subleading in the $\epsilon$-expansion. Solid lines represent fermion propagators while dotted lines represent propagators of complex bosons.}
\label{dgrmslrgeNc}
\end{figure}

\section{\label{sec:lrgeNc} $N_c \rightarrow \infty$ limit}
\begin{figure}[b]
\includegraphics[width=7.0cm, height=3.5cm]{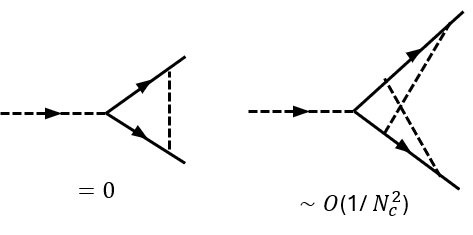}
\caption{a) One-loop vertex renormalization diagram, which turns out to be exactly zero. b) Two-loop vertex renormalization, which is of order $O(1/N^2_c)$.}
\label{vrtxrnmlztn}
\end{figure}

Only the boson field renormalization effect survives in the $N_c\rightarrow \infty$ limit (Fig. \ref{dgrmslrgeNc}a). The fermion renormalization is an $O(1/N_c)$ effect, hence suppressed in this limit (Fig. \ref{dgrmslrgeNc}b). The same is the case with four-boson interaction vertex renormalization by fermions, which is also an $O(1/N_c)$ effect (Fig. \ref{dgrmslrgeNc}b). While the Yukawa vertex is not renormalized at one-loop level, the two-loop contribution is an $O(1/N^2_c)$ effect(See Fig. \ref{vrtxrnmlztn}). 

After rescaling $\{\dfrac{g^2_{i}}{4\pi^2}, \dfrac{\lambda}{4\pi^2}  \}\rightarrow \{g^2_i,\lambda \}$, the RGE of the two couplings in the $N_c \rightarrow \infty$ limit becomes,
\begin{subequations}
\begin{eqnarray}
    \frac{dg_{i}}{dl} &=& g_{i} \left( \frac{\epsilon}{2}  - \frac{1}{2} \sum^{\mathcal{N}}_{j=1} g^2_j \right)\label{YkwaRGENcinfty1} \\
    \frac{d\lambda}{dl} &=& \lambda \left(\epsilon - 2 \sum^{\mathcal{N}}_{j=1} g^2_j  \right)   \label{YkwaRGENcinfty2} 
\end{eqnarray}    
\end{subequations}
Here, by increasing the scale $l$, the theory approaches the infrared limit. Fig. \ref{RGflowlrgeNcN2}(a) shows the renormalization group flow in the flavor space of an $\mathcal{N}=2$ flavor case. See appendix. \ref{appndx:RGE} for the derivation of the RGE in Eqs. (\ref{YkwaRGENcinfty1},\ref{YkwaRGENcinfty2})

Notice the emergent $SO(\mathcal{N})$ symmetry of the RGE in Eq. (\ref{YkwaRGENcinfty1}). If $\hat{R}$ is a rotation matrix in the $SO(\mathcal{N})$ group, then the RGE in Eq. (\ref{YkwaRGENcinfty1}) is invariant under the following transformation,
$$g_{i} \rightarrow R_{ij}g_j.$$

\subsection{$S^{\mathcal{N}-1}$ conformal manifold}

The solutions to Eq.\ref{YkwaRGENcinfty2} are evident, and we discuss them below.
These solutions explicitly indicate a conformal manifold, i.e. {\em a smooth manifold of scale-invariant Hamiltonians in the interaction parameter space}. By contrast, standard quantum critical phenomena are usually captured by well-isolated
Wilson-Fisher fixed points. So the phenomena here are quite distinct.


The strong-coupling fixed points found by solving for the zeroes of the beta function in Eq. (\ref{YkwaRGENcinfty1}) read,
\begin{eqnarray}
    \sum^{\mathcal{N}}_{i=1} g^2_{i,c} = \epsilon,\,\, \lambda_c = 0. \label{fxdpntsYkwalrgeNc}
\end{eqnarray}
Evidently, the solution describes an $(\mathcal{N}-1)$-sphere in the $\mathcal{N}$-dimensional interaction parameter space spanned by $\{g_i\}$, $i=1,...,\mathcal{N}$. Hence, we showed that a conformal manifold isomorphic to $S^{\mathcal{N}-1}$ exists, and it dictates the universality class of the phase boundary in the limit of $N_c \rightarrow \infty$. For $\mathcal{N}=2$, a ring conformal manifold is formed in the flavor space as shown in Fig. \ref{RGflowlrgeNcN2}(a).

The geometry of the conformal manifold can also be identified as a coset space $C = G/H$, where $G = SO(\mathcal{N})$ is the symmetry of the RGE and $H = SO(\mathcal{N}-1)$ is the invariant subgroup of the manifold. The standard group theory calculation indeed shows that $$ C = \frac{SO(\mathcal{N})}{SO(\mathcal{N}-1)} = S^{\mathcal{N}-1},$$
an ($\mathcal{N}-1$)-sphere in the flavor space.

\subsection{Anomalous dimensions}

Let us compute the anomalous dimensions $\eta_{\phi}$, $\eta_{\psi_i}$ of the bosonic and fermionic fields respectively in the $N_c \rightarrow \infty$ limit. It is given by,
\begin{subequations}
\begin{eqnarray}
    \eta_{\phi} &=& \sum^{\mathcal{N}}_{i=1} g^2_i \\
    \eta_{\psi_i} &=& 0.
\end{eqnarray}    
\end{subequations}

Fermion fields do not acquire an anomalous dimension because the renormalization of fermion fields is an  $O(1/N_c)$  effect. At the conformal manifold defined by Eq. (\ref{fxdpntsYkwalrgeNc}), the anomalous dimension of the boson field becomes, 
\begin{eqnarray}
 \eta_{\phi} = \epsilon   \label{anmlsbsnlrgeNc}
\end{eqnarray}
 This indicates that boson fields acquire an anomalous dimension due to their strong coupling with fermions, despite the weak interaction among themselves, as demonstrated by $\lambda_c = 0$.

\subsection{Deformation operators of a conformal manifold}

A conformal manifold implies that a conformal field theory defined at one point of the smooth manifold can be continuously deformed into another one in the near vicinity by applying deformation operators. For instance, let $\mathcal{S}_c$ be an action defined at a point on the conformal manifold and let $\mathcal{O}_{\text{D}}$ be the deformation operator. Then, after a deformation,
\begin{eqnarray}
    \mathcal{S}_c + \delta \mathcal{S} = \mathcal{S}_c + \delta \alpha \int d^{d}\textbf{x}\, d\tau \,\, \mathcal{O}_{\text{D}} (\textbf{x},\tau), 
\end{eqnarray}
the theory remains conformal invariant. Here $\delta \alpha$ is the corresponding coupling away from the fixed point.

A deformation operator, $\mathcal{O}_{\text{D}}$ (the subscript refers to deformation), lives in the tangent space of the conformal
manifold, and it shall be exactly marginal to be consistent with the manifold. The number of linearly independent tangential deformation operators shall define the dimension of the conformal manifold.

That is, these deformation operators shall not flow under the standard scale transformations, and the beta function assigned to those interaction operators shall be zero identically. Conformal manifolds
can be obtained by mapping out the smooth manifolds in which the beta functions become zero, i.e., {\em manifolds of zeros}. Exact marginal 
deformation operators at a particular location of the parameter space span a linear space that defines the tangent space of a manifold at that location.

Therefore, they shall have the scaling dimension exactly equal to the dimension of the space-time under consideration. For $d+1$ space-time dimensions, the deformation operator dimension shall exactly match the space-time dimension $d+1$,

\begin{eqnarray}
\Delta_{\mathcal{O}_{\text{D}}} =d+1. 
\end{eqnarray}

In contrast, at the free-fermion fixed point $g^2_c = 0$, for the same operator that appears in our studies, we find that
\begin{eqnarray}
    \Delta^{(0)}_{\mathcal{O}_{\text{D}}} = \frac{3 d}{2} - \frac{1}{2},
\end{eqnarray}
which is the engineering scaling dimensions of the operators. At $d=3$, the scaling dimensions at the two classes of fixed points match, not surprisingly, since the Yukawa interaction is marginal in this case. 

However, at $d=2$, the conformal manifold predicts a scaling dimension of $\Delta_{\mathcal{O}_{\text{D}}} =3$ while near the free fermion fixed point, the operator has a dimension of $2\frac{1}{2}$.

 For an $(\mathcal{N}-1)$-sphere conformal manifold, the number of marginal operators, $n_{MAR} = \mathcal{N}-1$. The exact marginal operators for $\mathcal{N}=2$ and $\mathcal{N}=3$ are shown in Eqs. (\ref{MgnloprtrN2}) and (\ref{Mgnloprtr1N3},\ref{Mgnloprtr2N3}).

 Eq. (\ref{anmlsbsnlrgeNc}) indicates that the anomalous dimensions of the bosons are the same throughout the conformal manifold.
 In addition, the scaling dimension of the Yukawa interaction operators ($O_{\text{Y}}$) [defined as $\mathcal{L}_{fb,N_c}$ in Eq. (\ref{YkwEFTNc})] at any point on the conformal manifold is given by $\triangle_{O_{\text{Y}}} = 4$ in the infinite $N_c$ limit.
 It is also independent of its location in the manifold. 
 Note that near a free fermion fixed point, these interaction operators instead have scaling dimension $\triangle^{(0)}_{O_{\text{Y}}} =2\frac{1}{2}$ (the superscript
 $0$ refers to the free fermion fixed point).

 Hence, the conformal manifold only hosts a single universality class. From a traditional quantum criticality point of view, we have named it a \textit{featureless} conformal manifold\cite{saran(2025)} to indicate that there is a unique set of critical indices. 
 
 However, the structure of these operators in the flavor space is different at different points on the manifold. We observed an EPR-like entanglement structure in the flavor space for the Yukawa interaction operators, which varies with the manifold coordinate. In Sec. (\ref{sec:entgnlmnt}), we quantify this difference by introducing a flavor space entanglement entropy. In Secs. (\ref{sec:N2Ykwa}) and (\ref{sec:N3Ykwa}), we calculate the entanglement entropy of the fixed point interaction operators for cases $\mathcal{N}=2$ and $\mathcal{N}=3$, and show that the entanglement structure is indeed a function of their respective manifold coordinates (See Figs. \ref{RGflowlrgeNcN2}(c) and \ref{EentrpyN3}).

\begin{figure}[b]
\includegraphics[width=8.6cm, height=8.cm]{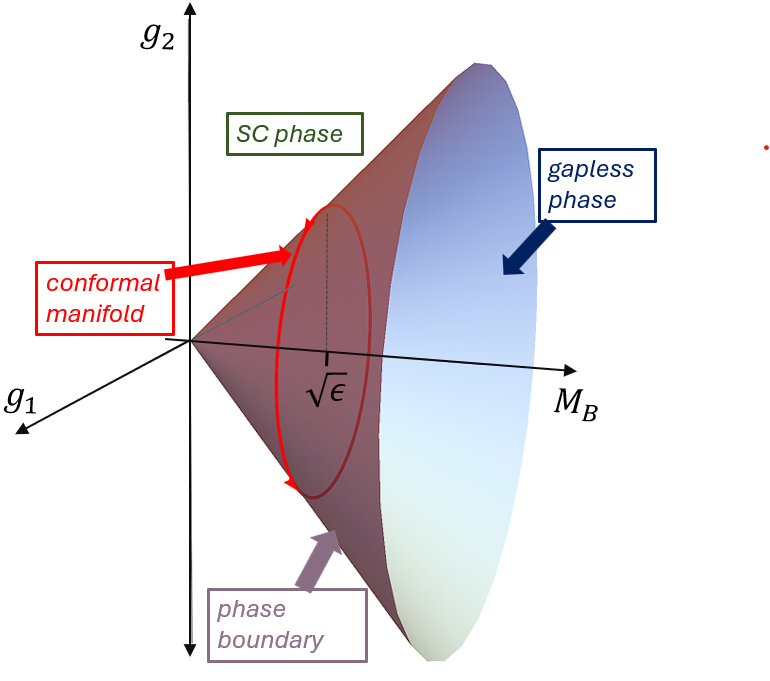}
\caption{Phase boundary of the gapless phase of $\mathcal{N}=2$ TI surface found using mean-field approximation (Eq. (\ref{phsebndrmnfldYkwa})) of the Yukawa theory in Eq. (\ref{YkwEFTNc}). The parameter space is spanned by the Yukawa couplings $(g_1,g_2)$ and the boson mass $M_B$. The red ring denotes the $S^1$ conformal manifold (Eq. (\ref{fxdpntsYkwalrgeNc})), which dictates the universality class of the phase boundary in the $N_c \rightarrow \infty$ limit. }
\label{phsebndry_ykwa}
\end{figure}

\subsection{Comparison with the mean-field theory} 
Here, we compare the phase boundary structure derived from the classical mean-field approximation with that from the large-$N_c$ renormalization group analysis. To facilitate this, we introduce the bare boson mass term as a perturbation to the massless theory in Eq. (\ref{YkwEFTNc}),
\begin{eqnarray}
\mathcal{S}_{Y,N_c}[\phi,\psi_i]&\rightarrow&  \mathcal{S}_{Y,N_c}[\phi,\psi_i] + \delta S^{M_B}_{Y,N_c}[\phi] \nonumber\\ 
   \text{where}\,\,\,  \delta S^{M_B}_{Y,N_c}[\phi] &=& \int d^d\textbf{r}d\tau \,\,    M_B^2 |\phi|^2 \nonumber 
\end{eqnarray}

At the mean-field level, we consider the bosonic field, $\phi$, constant and uniform. Integrating out the fermion degrees of freedom and varying the resultant effective action with respect to $|\phi|$ generates the mean-field equation. Taking the limit $|\phi| \rightarrow 0 $ in the gap equation, we arrive at the following equation for the phase boundary,
\begin{eqnarray}
    M^2_{B,p} = f(d) \sum^{\mathcal{N}}_{j=1} g^2_{j,p} \label{phsebndrmnfldYkwa}
\end{eqnarray}
where $f(d) = \frac{2 S_{d+1}}{(2\pi)^{(d+1)}} \frac{\Lambda^{d-1}}{d-1}$ and $\Lambda$ is the UV-cut-off of the interaction. Evidently, the phase boundary is a codimension-one manifold in $\mathcal{N}+1$-dimensional space spanned by $\{g_i\}$,$i=1,... \mathcal{N}$, and the mass parameter $M_B$. See Fig. \ref{phsebndry_ykwa} for the phase boundary in the $\mathcal{N}=2$ flavor case, where it is conical in shape.

Now we evaluate the critical boson mass by applying its renormalization group flow in the large-$N_c$ limit. The RG flow of the mass term is,
\begin{eqnarray}
    \frac{dM_B^2}{dl} &=& M_B^2 \left(2 - \sum_j g^2_j \right) - 2 \sum_{j} g^2_j
\end{eqnarray}
At the lowest order, the critical boson mass $M_B^2$ satisfies the relation,
\begin{subequations}
\begin{eqnarray}
    M_{B,c}^2 &=& \sum^{\mathcal{N}}_{j=1} g^2_{j,c} + O(g^4). \label{crtclmslrgeNc1} \\
    \implies && M_{B,c}^2 = \epsilon + O(\epsilon^2). \label{crtclmslrgeNc2}
\end{eqnarray} 
\end{subequations}

Here, both the couplings are dimensionless. 
Evidently, the equation for critical mass in Eq. (\ref{crtclmslrgeNc1}) is the same as the mean-field result in Eq. (\ref{phsebndrmnfldYkwa}). However, the similarity ends here. While the boson mass is not constrained to a specific value in the mean-field approximation, as indicated by Eq. (\ref{phsebndrmnfldYkwa}), the large-$N_c$ renormalization group analysis implies that the theory remains scale-invariant only if  $M_B^2$ is further fine-tuned to a critical value $M_{B,c}^2$ defined in Eq.\ref{crtclmslrgeNc2} which only depends on the dimensionality of $d=3-\epsilon$

Therefore, in the $N_c \rightarrow \infty$ limit, the phase boundary is simply an $\mathcal{N}$-dimensional manifold (Eq. (\ref{phsebndrmnfldYkwa})) in the $\mathcal{N}+1$-dimensional parameter space spanned by $\{g_i\}$ and $m$, where $i=1,.,\mathcal{N}$. 

On the other hand, the conformal manifold is a $\mathcal{N}$-submanifold isomorphic to $S^{\mathcal{N}-1}$, further embedded in the phase boundary. It is a co-dimension two manifold in the $\mathcal{N}+1$-dimensional parameter space spanned by $\{g_i\}$ and $m$, where $i=1,.,\mathcal{N}$. This manifold (Eq. (\ref{fxdpntsYkwalrgeNc})) dictates the universality class of the $\mathcal{N}$-dimensional phase boundary.

\section{\label{sec:entgnlmnt}Entanglement}

In this section, we derive the flavor-space entanglement entropy of the fermion-boson interaction operators of the action in Eq. (\ref{YkwEFTNc}), which have a Yukawa structure. 

Define a generalized Yukawa operator as,
\begin{eqnarray}
    \mathcal{O}_{\text{Y}} = \phi^{*}\sum^{\mathcal{N}}_{i,j=1}  A_{ij} \psi_i^{T}(is_y)\psi_j + \text{h.c} \label{Ykwaoprtr}
\end{eqnarray}
where $A_{ij}$ is an $\mathcal{N} \times \mathcal{N}$ matrix defining the structure of the operators. Note that $A_{ij}$ can be complex matrices in general, although in our studies we restrict to real matrices to impose the time reversal symmetry. 

The trilinaer operator in Eq. (\ref{Ykwaoprtr}) is also maximally entangled in the fermion spin subspace. However, the entanglement entropy associated with that remains uniform throughout the conformal manifold and won't differentiate theories at different points. (see Appendix \ref{appndxEentrpy}). Hence, we mute the constant spin entropy for the discussions below.  

We can introduce a flavor space reduced density matrix as,
\begin{eqnarray}
    \rho_F = \frac{1}{\text{Tr}[AA^{\dagger}]} AA^{\dagger}, \,\, \text{Tr}[\rho_F] =1
\end{eqnarray}
where the sub-index $F$ refers to the flavor space.

$\rho_{F}$ defined in this way is an $\mathcal{N}\times \mathcal{N}$ Hermitian matrix that is bilinear in terms of operator matrix $A_{ij}$. It is obtained by tracing over one of the flavor indices $i,j=1,..., \mathcal{N}$ in the product of $A^\dagger A$. 

A normalization constant has been introduced to ensure that the trace identity of the density matrix holds. 

The entanglement entropy defined in terms of $\rho_{F}$ is,
\begin{eqnarray}
    S_{F} = - \text{Tr}\left[\rho_{F} \ln \rho_{F} \right]
\end{eqnarray}

For our discussions of the Yukawa interactions, the $A_{ij}$ has a very simple structure of $A_{ij} = a_i \delta_{ij}$. Therefore, 
\begin{eqnarray}
    S_{F} = - \sum^{\mathcal{N}}_{i=1} a^2_{i} \ln a^2_i.\label{Eentrpy}
\end{eqnarray}
$S_{F}$ is naturally a function of point $p$ on the conformal manifolds that emerge as exact solutions to RGEs at $N_c \rightarrow \infty$. Note that, as expected, $S_F$ is bounded from above by $\ln \mathcal{N}$.

That is, every point p on the manifold represents a distinct scale-conformal invariant Hamiltonian in the infinite $N_c$ limit. Note that all the conformal manifolds are simply connected, isomorphic to $S^{\mathcal{N}-1}$.

\begin{figure*}
    \centering
    \includegraphics[width=7.5cm, height=7.cm]{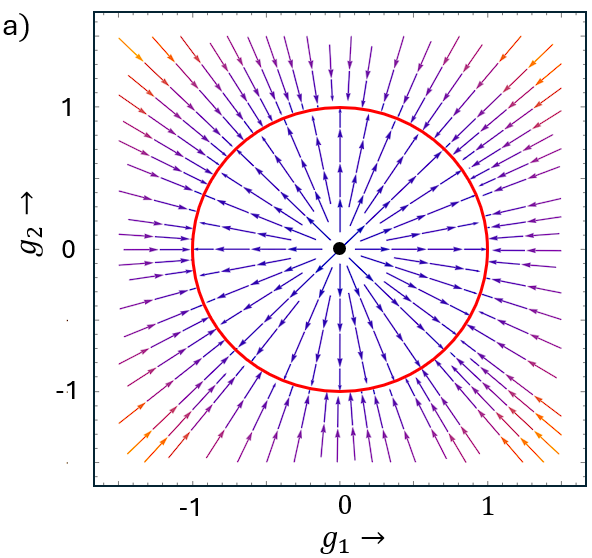}
    \includegraphics[width=7.5cm, height=7.14cm]{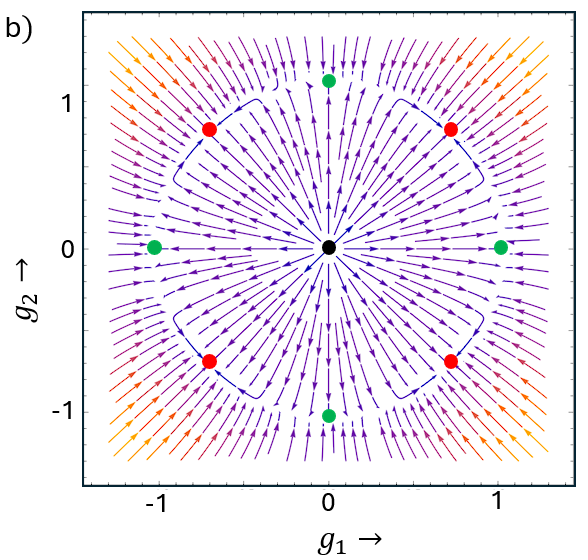}
    \includegraphics[width=9.5cm, height=6.18cm]{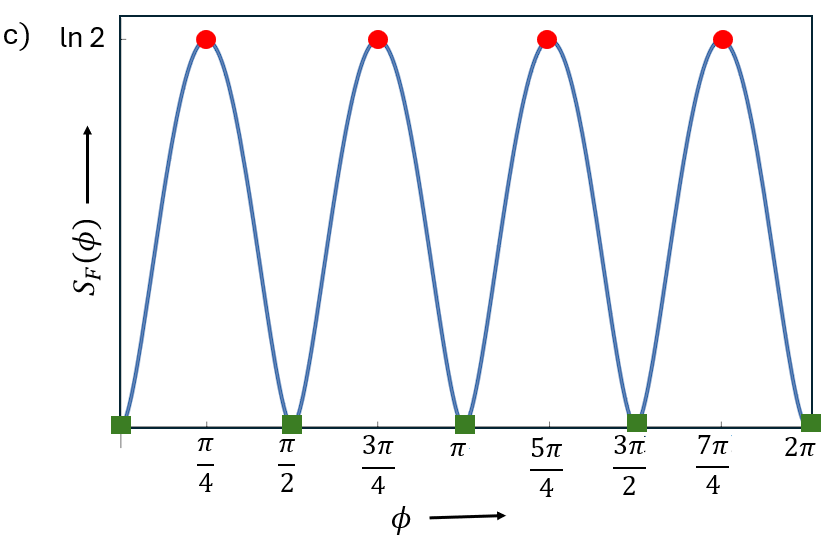}
    
    \caption{(a) Renormalization group (RG) flow in the plane of the Yukawa couplings $(g_1,g_2)$ for $\mathcal{N}=2$ in the limit of $N_c \rightarrow \infty$ (Eqs. (\ref{YkwaRGENcfnte2N1},\ref{YkwaRGENcfnte2N2})). The red-colored ring describes the conformal manifold (Eq. (\ref{cnfmlmnfld_N2})), i.e., a smooth manifold of scale-invariant fixed-point Hamiltonians, while the black dot is the isolated free-fermion fixed point. 
    The directions of flow are towards the IR scales.
    (b) RG flow studied at first order in $1/N_c$ and at two-loop level. The conformal manifold in (a) breaks into isolated fixed points.  Red dots represent the IR stable universality class (class I, Eq. (\ref{IRstbleN2})), while the green dots describe the IR unstable set of fixed points(class II, Eq. (\ref{IRunstbleN2}).  We set $\epsilon = 0.95$ and $N_c = 10$.
    (c) Entanglement entropy $S_F(\phi)$ of the fixed point interaction operator evaluated along the $S^1$ manifold (see panel (a) ), plotted as a function of the angular coordinate $\phi$ (Eq. (\ref{EentrpyN2})) in the limit of $N_c \rightarrow \infty$. The red circles and the green squares correspond to the isolated fixed points from (b), appearing at first order in $1/N_c$.}
    \label{RGflowlrgeNcN2}
\end{figure*}

In Figs. \ref{RGflowlrgeNcN2}(c) and \ref{EentrpyN3}, we show explicitly $S_{F}$ as a function of $p$ point on the conformal manifolds for the case of $\mathcal{N}=2$ and $\mathcal{N}=3$ respectively.

 \section{Quantum fluctuations of the order of $O(1/N_c)$ }
Here, we compute the conformal manifold breaking terms in the RGE that appear at first order in $1/N_c$. We compute both the one-loop and two-loop renormalization effects that are also first order in $1/N_c$ (Fig. \ref{dgrmslrgeNc}). 

At the one-loop level, the renormalization of the fermion field and the four-boson interaction vertex constitute the $O(1/N_c)$ contributions (Fig. \ref{dgrmslrgeNc}(b)). 

At the two-loop level, $O(1/N_c)$ effects include the two sunrise-type diagrams contributing to the boson and the fermion field renormalization effects (Fig. \ref{dgrmslrgeNc}c). 

As illustrated in Fig. \ref{vrtxrnmlztn}, the renormalization of the Yukawa vertex begins to appear only at the order of \( O(1/N^2_c) \).

Putting together, the beta function of the dimensionless couplings $g^2_i$ and $\lambda$ reads (see appendix. \ref{appndx:RGE} for the derivation),
\begin{subequations}
\begin{eqnarray}
   \frac{d g_i}{dl} &=& \frac{\epsilon}{2}  g_i - \frac{1}{2} g_{i} \sum^{\mathcal{N}}_{j=1} g^2_j - \frac{g^3_{i}}{N_c}\nonumber \\ &+& \frac{3}{4} \frac{g_i}{N_c}  \sum^{\mathcal{N}}_{j=1} g^4_j + \frac{3}{4} \frac{g^3_i}{N_c} \sum^{\mathcal{N}}_{j=1} g^2_{j}\label{YkwaRGENcfntegen1}  \\
   \frac{d\lambda}{dl} &=& \epsilon \lambda - 2\lambda  \sum^{\mathcal{N}}_{j=1} g^2_j + \frac{4}{N_c} \sum^{\mathcal{N}}_{j=1} g^4_j \label{YkwaRGENcfntegen2}
\end{eqnarray}\
\end{subequations}
The anomalous dimensions of the bosonic and the fermionic fields read,
\begin{subequations}
\begin{eqnarray}
    \eta_{\phi} &=& \sum^{\mathcal{N}}_{j=1} g^2_j - \frac{3}{2} \sum^{\mathcal{N}}_{j=1} \frac{g^4_j}{N_c} \label{anmlsfnteNcbsn} \\
    \eta_{\psi_i} &=&  \frac{g^2_i}{N_c} - \frac{3}{4} \frac{g^2_i}{N_c} \sum^{\mathcal{N}}_{j=1} g^2_j \label{anmlsfnteNcfrmn}
\end{eqnarray}
\end{subequations}

Notice in Eq. (\ref{anmlsfnteNcfrmn}) that the fermions of different flavors acquire anomalous dimensions at order $O(1/N_c)$, even at one-loop level. As a result, the RGE in Eq. (\ref{YkwaRGENcfntegen1}) is no longer invariant under the $SO(\mathcal{N})$ rotations of the couplings $g_i$.

As expected, the $O(1/N_c)$ contributions break the emergent $SO(\mathcal{N})$ symmetry of the RGE. Consequently, the conformal manifolds observed in the large-$N_c$ limit break down into isolated fixed points. In the coming sections, we will study the properties of these isolated fixed points for the TI surfaces with $\mathcal{N}=2,3$ fermion flavors. 

\subsection{\label{secRGEgensubsecsmmtry}Discrete symmetries of the RGE} Even though the RGE does not possess the continuous $SO(\mathcal{N})$ symmetry, it does have some discrete symmetries in the flavor space. These discrete symmetries continue to play a significant role in classifying the fixed points and, consequently, in identifying the universality classes of the theory. For instance, if $G$ is the symmetry group of the RGE and $H$ is the invariant subgroup of the fixed point(s), then the set of fixed points can be identified as the coset space $C = G/H$ \cite{saran(2025)}.  All the fixed points in this coset space belong to the same universality class.

We now examine the discrete transformations that leave the RGE in Eq. (\ref{YkwaRGENcfntegen1}) invariant.
It is invariant under individual sign flips of each coupling constant: $g_i \rightarrow -g_i$, independently for each $i$. This generates a symmetry group of the form $\left(Z_{2}\right)^{\mathcal{N}}$. Secondly, if we represent the set of couplings,$\{g_{i}\}$ as a discrete set of $\mathcal{N}$ elements, the permutation group  $S_{\mathcal{N}}$ that permutes the flavor indices is also a symmetry of the RGE. Overall, the symmetry group, $G$, of the RGE in Eq. (\ref{YkwaRGENcfntegen1}) is,
\begin{eqnarray}
    G = S_{\mathcal{N}} \otimes \left( Z_{2} \right)^{\mathcal{N}}.\label{smmtrygrpRGEgen}
\end{eqnarray}
    
\section{\label{sec:N2Ykwa}Main result I:\\ Conformal manifolds and Wilson-Fisher fixed points for $\mathcal{N}=2$ flavor}

In this section, we focus on the $\mathcal{N}=2$ flavor TI surface. The flavor space is 2-dimensional. First, we study the properties of the conformal manifold that emerges in the large-$N_c$ limit (see Sec.\ref{sec:lrgeNc}). Next, we examine the renormalization group (RG) flow along this ring manifold, which begins to appear at first order in $1/N_c$, as dictated by Eqs. (\ref{YkwaRGENcfntegen1},\ref{YkwaRGENcfntegen2}).

Since the conformal manifold is isomorphic to $S^1$, we shall use polar coordinates for the couplings $(g_1,g_2)$ in the flavor space,
\begin{eqnarray}
    g_1 = g\cos \phi,\,\, g_2 = g \sin \phi, \,\, \phi \in [0,2\pi].\nonumber
\end{eqnarray}
Setting $\mathcal{N}=2$ and in terms of the polar coordinates, $g$ and $\theta$, the RGE in Eqs. (\ref{YkwaRGENcfntegen1},\ref{YkwaRGENcfntegen2}) become,
\begin{subequations}
    \begin{eqnarray}
    \frac{d g}{d l} &=& \frac{\epsilon}{2} g - \frac{1}{2} g^3 \nonumber \\ &-&  \left(1 - \frac{1}{2} \sin^2 2\phi \right) \left[ \frac{g^3}{N_c} - \frac{3g^5}{2N_c} \right]\label{YkwaRGENcfnte2N1} \\
    \frac{d\phi}{dl} &=& \frac{g^2}{4N_c} \sin 4\phi \left(1 - \frac{3}{4} g^2 \right)  \label{YkwaRGENcfnte2N2} \\
    \frac{d\lambda}{dl} &=& \epsilon \lambda - 2\lambda g^2 + \frac{4 g^4}{N_c}  \left(1 - \frac{1}{2} \sin^2 2\phi \right) \label{YkwaRGENcfnte2N3}
\end{eqnarray}
\end{subequations}

\subsection{$S^1$ Conformal manifold in the $N_c \rightarrow \infty$}
In the $N_c\rightarrow \infty$ limit, the flow along the azimuthal direction described by Eq. (\ref{YkwaRGENcfnte2N2}) vanishes for all $\phi \in [0,2\pi]$. Hence, a conformal manifold in the shape of $S^{1}$ exists in the flavor space described by the equation,
\begin{eqnarray}
    g_c = \sqrt{\epsilon}. \label{cnfmlmnfld_N2}
\end{eqnarray}
Fig. \ref{RGflowlrgeNcN2}(a) shows the renormalization group flow in the flavor space.

As discussed in Sec. (\ref{sec:lrgeNc}), the conformal manifold is characterized by exactly marginal deformation operators. At a point $(g_{1,c}, g_{2,c}) = (\sqrt{\epsilon}\cos \phi, \sqrt{\epsilon}\sin \phi)$ on the ring manifold, the deformation operator $\mathcal{O}_{\text{D}}(\phi)$ and its scaling dimension $\Delta_{\mathcal{O}_{D}}$ is given by,
\begin{eqnarray}
    \mathcal{O}_{\text{D}}(\phi)  &=&  \phi^* \bigl[- \sin \phi \,\, \psi^{T}_{1}\left(i s_y \right) \psi_{1}  + \cos \phi \,\, \psi^{T}_{2}\left(i s_y \right) \psi_{2} \bigr]\nonumber \\ &+& \text{h.c.} \label{MgnloprtrN2} \\
    \Delta_{\mathcal{O}_{\text{D}}} &=& d+1. \nonumber
\end{eqnarray}

Furthermore, we can also study the variations of the Yukawa interaction operators themselves as we move around the conformal manifold. As a function of the azimuthal angle $\phi$, the operator and its scaling dimension are given by,
\begin{eqnarray}
    \mathcal{O}_{\text{Y}}(\phi) &=& \phi^* \left[  \cos \phi \,\, \psi^{T}_{1}\left(i s_y \right) \psi_{1}  + \sin \phi \,\, \psi^{T}_{2}\left(i s_y \right) \psi_{2} \right] \nonumber \\ &+& \text{h.c}
    \\  \label{RdloprtrN2}  \Delta_{\mathcal{O}_{\text{Y}}} &=& 4.\nonumber
\end{eqnarray}

Notice that both the marginal deformation and the Yukawa interaction operators at the manifold are functions of locations $p$ on the manifold, parametrized by the azimuthal angle $\phi$. We observe an EPR-like entanglement structure in both these operators in the flavor space, which varies with $\phi$ on the manifold.

In Sec. (\ref{sec:entgnlmnt}), we developed a formula to evaluate the entanglement entropy of the Yukawa-like operators in the flavor space (Eq. (\ref{Eentrpy})).
Here, we use that formula to calculate the entanglement entropy of the interaction operator in Eq. (\ref{RdloprtrN2}).
The entanglement entropy of the fixed point interaction operator has the form,
\begin{eqnarray}
    S_{F}(\phi) = - \left[ \sin^2 \phi \ln \sin^2 \phi + \cos^2 \phi \ln \cos^2\phi   \right]\label{EentrpyN2}
\end{eqnarray}
In Fig.\ref{RGflowlrgeNcN2}(c), we plot $S_{F}$ as a function of the azimuthal angle $\phi$.
Eq. (\ref{EentrpyN2}) reveals that the operator is maximally entangled at points, $$\phi_{2k-1} = (2k-1)\pi/4,\,\,k=1,2,3,4,$$  on the manifold. For example, at $\phi_1 = \pi/4$, the interaction operator and the entanglement entropy read,
\begin{eqnarray}
    \mathcal{O}_{\text{Y}}(\frac{\pi}{4}) &=& \frac{\phi^{*}}{\sqrt{2}}  \left[\psi^{T}_{1}\left(i s_y \right) \psi_{1} + \psi^{T}_{2}\left(i s_y \right) \psi_{2} \right]  + \text{h.c} \label{N2lrgeNcentngled} \\
    S_{F} (\frac{\pi}{4}) &=& \ln 2.\nonumber
\end{eqnarray}
The operator is maximally entangled in the flavor space as expected.

On the other hand, the entanglement entropy, $S_{F} = 0$, at points $$\phi_{2k} = k\pi/2,\,\,k=1,2,3,4.$$ For example, at $\phi = 0$, we have,
\begin{eqnarray}
    \mathcal{O}_{\text{Y}}(0) &=& \psi^{T}_{1}\left(i s_y \right) \psi_{1} \phi^{*} + \text{h.c.} \label{N2lrgeNcprdct} \\
    S_{F}(0) &=& 0 \nonumber
\end{eqnarray}
In this case, the interaction operator is in a product state of the fermion field operators in flavor space, resulting in an entanglement entropy of zero.

The entanglement entropy of the exactly marginal operator also has the same form as in Eq. (\ref{EentrpyN2}). Thus, we demonstrated that the entanglement structure of the operators, as quantified by the entanglement entropy, varies as a function of the manifold coordinate. 

\begin{table*}
 \caption{Isolated fixed points, anomalous dimensions of bosonic (Eq. (\ref{anmlsfnteNcbsn})) and fermionic fields (Eq. (\ref{anmlsfnteNcfrmn})) expanded to order $O(\epsilon^2,1/N_c)$ of an $\mathcal{N}=2$ flavor TI surface. Isolated fixed points obtained from solving the RGEs in Eqs. (\ref{YkwaRGENcfnte2N1},\ref{YkwaRGENcfnte2N2}). Here $\phi \in [0,2\pi]$ and $(i,j) = \{1,2\}$. $n_{IRR}$ and $n_R$ stand for the number of irrelevant and relevant operators, respectively. $S_{n}$ stands for permutation group of size $n$.}
    \label{ykwaN2tble}
\renewcommand{\arraystretch}{4.5}
\begin{tabular}{|c|c|c|c|c|c|c|}
\hline\hline
    \makecell{\,\,\,} &
    \multicolumn{2}{|c|}{Fixed points} &
    \makecell{Number of\\ relevant / \\ irrelevant  \\ operators}  & 
    \makecell{Anomalous dimensions} &
    \makecell{Coset space} & \makecell{IR stability} 
    \\ \cline{1-2}
    \hline
       I &
    \makecell{$\phi_c = (2n+1) \dfrac{\pi}{4}$ \\[0.9em] $n\in\{0,1,2,3 \}$} & 
     \makecell[l]{  $g^2_c = \epsilon - \dfrac{\epsilon}{N_c} + \dfrac{3}{2} \dfrac{\epsilon^2}{N_c}$ \\[0.7em] $\lambda_c = \dfrac{2\epsilon}{N_c}$}
     &
    \makecell{$n_{IRR} = 2, $\\$n_{R} = 0$}    & 

     \makecell[l]{$\eta_{\phi} = \epsilon - \dfrac{\epsilon}{N_c} + \dfrac{3}{4} \dfrac{\epsilon^2}{N_c}$ \\[0.7em]$
    \eta_{\psi_i}= \dfrac{\epsilon}{2N_c} - \dfrac{3}{8}\dfrac{\epsilon^2}{N_c}$, \\[0.9em]$ i\in\{1,2\}$}
       &  
       \makecell{$Z_2 \otimes Z_2$} 
       &
         IR stable
          \\ 
     \hline

       II  &
     \makecell[l]{$\phi_c = \dfrac{n\pi}{2}$ \\[0.9em] $n\in\{0,1,2,3 \}$} &

     \makecell[l]{ $g^2_c = \epsilon - \dfrac{2\epsilon}{N_c} + \dfrac{3\epsilon^2}{N_c}$ \\[0.7em] $\lambda_c = \dfrac{4\epsilon}{N_c} $}
     
      &  \makecell{$n_{IRR} = 1,$\\$n_{R} = 1$ }    &   
      \makecell[l]{ $\eta_{\phi} = \epsilon - \dfrac{2\epsilon}{N_c} + \dfrac{3}{2} \dfrac{\epsilon^2}{N_c}$ \\[0.7em] $\eta_{\psi_i} = \dfrac{\epsilon}{N_c} - \dfrac{3}{4} \dfrac{\epsilon^2}{N_c} $  \\[0.7em]  $\eta_{\psi_{j\neq i}} =0$}   
                 &
        \makecell{$S_2 \otimes Z_2$} 
        &
        IR unstable
        
        \\ \hline \hline
    \end{tabular}
    
\end{table*}

\subsection{Isolated Wilson-Fisher fixed points at $O(1/N_c)$}

At first order in $1/N_c$,
Eq. (\ref{YkwaRGENcfnte2N2}) indicates that a finite flow is induced along the ring conformal manifold. The effects of fermion field renormalization at both one- and two-loop levels contribute to this flow in this limit. However, the two-loop boson renormalization effect, shown in Fig. \ref{dgrmslrgeNc}(e), does not induce a flow along the manifold, even though it breaks the $SO(2)$ symmetry of the RGE. 

As a result, the conformal manifold breaks down into isolated Wilson-Fisher-like fixed points in the flavor space, as Fig. \ref{RGflowlrgeNcN2}(b) shows. The flow along the azimuthal direction, given by Eq. (\ref{YkwaRGENcfnte2N2}), vanishes only at specific discrete points, $\phi_c = n\pi/4,\,\, n = 0,1,...,7$. Thus, the conformal manifold breaks down into $3^2 - 1 = 8$ distinct isolated fixed points. Fig.\ref{RGflowlrgeNcN2}(b) shows the renormalization group flow in the flavor space. Based on the discrete symmetry arguments made in Sec. (\ref{secRGEgensubsecsmmtry}), we classify them into two distinct universality classes, namely,
\begin{itemize}
    \item Class I: \begin{equation}
        \phi_c = (2n+1) \pi/4, \,\, (n=0,1,2,3 ) \label{IRstbleN2}
    \end{equation}.
    \item Class II:
     \begin{eqnarray}
         \phi_c = n\pi/2,\,\, (n=0,1,2,3 ) \label{IRunstbleN2}
     \end{eqnarray} . 
\end{itemize}
We study their properties below:

\paragraph*{\textbf{Class I}:} In this universality class, both the fermion flavors are strongly coupled to the bosons at an equal magnitude (red dots in Fig.\ref{RGflowlrgeNcN2}(b)). Yukawa coupling strength expanded to $O(\epsilon^2,1/N_c)$ in cartesian coordinates reads,
\begin{eqnarray}
  g^2_{1,c} = g^2_{2,c} = \frac{\epsilon}{2} - \dfrac{\epsilon}{2N_c} + \dfrac{3}{4} \dfrac{\epsilon^2}{N_c} \label{N2fxpnts1}  
\end{eqnarray}

  The invariant subgroup of the fixed points is $H = S_2$. For $\mathcal{N}=2$, Eq. (\ref{smmtrygrpRGEgen}) suggests that the symmetry group of the RGE in Eq. (\ref{YkwaRGENcfntegen1}) is $G = S_{2} \otimes (Z_{2})^2$. Thus, the set of fixed points forms elements of the coset space, $$C = \frac{S_{2} \otimes (Z_{2})^2}{S_2}= Z_2 \otimes Z_2.$$ A total of $N(C) = 4$ fixed points belong to this universality class.  The table \ref {ykwaN2tble} lists the fixed points $ g^2_c$ and $ \lambda_c$, as well as the anomalous dimensions of the fermionic and bosonic fields, expanded to the order $O(\epsilon^2, 1/N_c)$.
  
  This specific set of fixed points, defined by Eq. (\ref{IRstbleN2}), turns out to be infrared stable (see Fig. \ref{RGflowlrgeNcN2}(b)). The deformation operators, defined in Eq. (\ref{MgnloprtrN2}) at these fixed points (Eq. (\ref{IRstbleN2})), which were exactly marginal in the $ N_c\rightarrow\infty$ limit, have now become weakly irrelevant at first order in $1/N_c$.

  Notably, the Yukawa interaction operator at these fixed points, obtained by plugging Eq. (\ref{IRstbleN2}) into Eq. (\ref{RdloprtrN2}), turns out to be maximally entangled in the flavor space, as shown in the table. \ref{YkwaN2dfmtn} (also pointed out as red circles in Fig. \ref{RGflowlrgeNcN2}(c)). 

 Therefore, we observe that the quantum fluctuations at $O(1/N_c)$ result in a renormalization group flow towards those points where the fixed point interaction operators are maximally entangled (Figs. \ref{RGflowlrgeNcN2}(b,c)).

\paragraph*{\textbf{Class II}:}  Only one of the two fermion flavors couples with the bosons in this universality class (green dots in Fig.\ref{RGflowlrgeNcN2}(b)). We have,
\begin{eqnarray}
  g^2_{i,c} = 0,\,\,g^2_{j,c} =\epsilon - \dfrac{2\epsilon}{N_c} + \dfrac{3\epsilon^2}{N_c}  \label{N2fxpnts2}
\end{eqnarray}
 for distinct $i,j \in \{ 1,2\}$ (see Table (\ref{ykwaN2tble}) for detailed solutions).
 
 The invariant subgroup of the fixed points is $H = Z_2$, hence the coset space $$C = S_2 \otimes Z_2.$$ We have a total $N(C) = 4$ fixed points in this class. Since the Yukawa coupling strength of the other fermion flavor needs to be fine-tuned to zero, this set of fixed points is IR unstable  (see Fig.\ref{RGflowlrgeNcN2}(b)). Table \ref{ykwaN2tble} summarizes the results.

The deformation operator, defined in Eq. (\ref{MgnloprtrN2}) at these fixed points (Eq. (\ref{IRunstbleN2})) has become weakly relevant at finite $ N_c$. Furthermore, the entanglement entropy ($S_F$) of the Yukawa interaction operator (Eq. (\ref{RdloprtrN2})) at these fixed points (Eq. \ref{IRunstbleN2}) is zero, as indicated in the table. \ref{YkwaN2dfmtn} (shown as green squares in Fig. \ref{RGflowlrgeNcN2}(c)).

 We want to point out further that these IR unstable fixed points are equivalent to SUSY fixed points previously studied in Ref.\cite{Ponte(2014),Grover(2014),maciejko(2016),yao(2017),yao(2017)(2),yao(2018)} for the case of $\mathcal{N}=1$ with a single two-dimensional Dirac cone. In our studies here with $\mathcal{N}=2$ and $\mathcal{N}=3$ 
(see below), Such SUSY fixed points still appear as a subset of fixed points (noted as Class II here, green squares in Fig. \ref{RGflowlrgeNcN2}(c)). And the corresponding flavor space entanglement entropy vanishes when approaching these fixed points (green squares in Fig. \ref{RGflowlrgeNcN2}(c)).

However, these SUSY fixed points are unstable and do not define the physics of surface topological quantum criticality or universality classes in the limit of interest. The IR stable fixed points, which are maximally entangled and dictate surface topological quantum criticality, do not exhibit SUSY features, unlike in the case of a single surface Dirac cone. This is one of the distinct features of our results when compared with previous studies.

In summary, for the $\mathcal{N}=2$ flavor TI surface in the $N_c \rightarrow \infty$ limit, the universality class is governed by a conformal manifold isomorphic to $S^1$. The flavor space entanglement entropy of the interaction operator varies with the manifold coordinate. Finite-$N_c$ quantum fluctuations generate a flow along this manifold, breaking it into isolated Wilson–Fisher-like fixed points belonging to two distinct universality classes. The class I, which is infrared stable, is characterized by maximally entangled interaction operators in the flavor space (see Table \ref{YkwaN2dfmtn}) and dictates the surface topological quantum criticality(sTQCP). On the other hand, the SUSY fixed points discussed above (class II) correspond to interaction operators with zero entanglement entropy and are IR unstable.  In other words, finite-$N_c$ RG flows along the manifold toward fixed points where the interaction operators are maximally entangled in flavor space.


\begin{table}[H]
 \caption{The Yukawa interaction operators at selected points in the two universality classes of an $\mathcal{N}=2$ flavor TI surface at first order in $1/N_c$. We list their scaling dimensions, expanded to $O(\epsilon^2, 1/N_c)$ and the entanglement entropy, $S_F$ (Eq. (\ref{EentrpyN2})). } 
 \label{YkwaN2dfmtn} 
   \renewcommand{\arraystretch}{3.0}
\begin{tabular}{|c|c|c|c|c|}
\hline \hline 
\makecell{} & 
\makecell{ Fixed \\ point }& \makecell{Yukawa\\ interaction operator}  & \makecell{Scaling \\ dimens-\\ions} & \makecell{Entanglement\\ entropy \\($S_F$)}
 \\ \hline
 I  &
\makecell{ $\phi_c = \dfrac{\pi}{4}$} & 
 
 \makecell{$\dfrac{\phi^{*}}{\sqrt{2}}  [\psi^{T}_{1}\left(i s_y \right) \psi_{1} +$ \\[1.0em]  $\psi^{T}_{2}\left(i s_y \right) \psi_{2} \bigr]$ + h.c }

 & \makecell{$4 - \dfrac{3 \epsilon^2}{2N_c}$}  & $\ln 2$
 
 \\  \hline

 II &
 
 \makecell{$\phi_c = 0$} & 
 
 \makecell{$\psi^{T}_{1}\left(i s_y \right) \psi_{1} \phi^{*} + \text{h.c.}$} &
 
 $4 - \dfrac{3\epsilon^2}{N_c}$ & $0$
   \\ \hline \hline
 \end{tabular}
\end{table}

\section{\label{sec:N3Ykwa}Main result II:\\ Conformal manifolds and Wilson-Fisher fixed points for $\mathcal{N}=3$ flavor}

For an $\mathcal{N}=3$ flavor TI surface, the flavor space is 3-dimensional. In the $N_c \rightarrow \infty$ limit, we saw in Sec. (\ref{sec:lrgeNc}) that a conformal manifold isomorphic to $S^2$ dictates the universality class of the phase boundary. In this section, we first examine the properties of the conformal manifold, especially the entanglement structure of its operators in the flavor space. Then, we investigate the finite-$N_c$ effects that result in a renormalization group flow along the manifold, as in Eqs. (\ref{YkwaRGENcfntegen1},\ref{YkwaRGENcfntegen2}), to first order in $1/N_c$.

Since the conformal manifold has spherical symmetry, it is convenient to shift to spherical coordinates in the flavor space. Define,
\begin{eqnarray}
    g_1 = g \sin\theta \cos \phi, \,\, g_2 = g \sin\theta \sin \phi, \,\, g_3 = g \cos \theta.\nonumber
\end{eqnarray}
In terms of the spherical coordinates, the RGE in Eqs. (\ref{YkwaRGENcfntegen1},\ref{YkwaRGENcfntegen2}) become,
\begin{subequations}
\begin{eqnarray}
    \frac{dg}{dl} &=& \frac{\epsilon}{2} g - \frac{1}{2} g^3 - f_g \left(\theta, \phi \right) \left[ \frac{g^3}{N_c} - \frac{3}{2} \frac{g^5}{N_c} \right]\label{YkwaRGENcfnte3N1} \\
    \frac{d\theta}{dl} &=&  \frac{g^2}{2N_c} \left(1 - \frac{3}{4} g^2 \right) f_{\theta} (\theta, \phi) \label{YkwaRGENcfnte3N2} \\
     \frac{d \phi}{d l} &=&  \frac{g^2}{4N_c} \left( 1 - \frac{3}{4} g^2\right) f_{\phi} (\theta, \phi) \label{YkwaRGENcfnte3N3} \\
     \frac{d\lambda}{dl} &=& \epsilon \lambda - 2\lambda g^2 + \frac{4\, g^4}{N_c} f_{\lambda} (\theta, \phi) \label{YkwaRGENcfnte3N4}
\end{eqnarray}    
\end{subequations}
where
\begin{eqnarray*}
   f_{g}(\theta, \phi) &=& f_{\lambda}(\theta, \phi) = 1 - \frac{1}{2}\sin^2 2\theta - \frac{1}{2} \sin^4 \theta \sin^2 2\phi \\
   f_{\theta}(\theta, \phi) &=& \left( \cos 2\theta + \frac{1}{2} \sin^2\theta \sin^22\phi \right) \sin 2\theta \\
    f_{\phi}(\theta, \phi) &=& \sin 4\phi \sin^3 \theta.
\end{eqnarray*}

\begin{table*}
 \caption{Isolated fixed points, anomalous dimensions of bosonic (Eq. (\ref{anmlsfnteNcbsn})) and fermionic fields (Eq. (\ref{anmlsfnteNcfrmn})) expanded to order $O(\epsilon^2,1/N_c)$ of an $\mathcal{N}=3$ flavor TI surface. Isolated fixed points obtained from solving the RGEs in Eqs. (\ref{YkwaRGENcfnte3N1},\ref{YkwaRGENcfnte3N2},\ref{YkwaRGENcfnte3N3}). Here $\phi \in [0,2\pi]$ and $i,j = 1,2.3$. $n_{IRR}$ and $n_R$ stand for the number of irrelevant and relevant operators, respectively. $C_{n}$ stands for cyclic group of size $n$. }
    \label{ykwaN3tble}
\renewcommand{\arraystretch}{4.5}
\begin{tabular}{|c|c|c|c|c|c|c|}
\hline\hline  
    \makecell{\,\,\,\,\,\,\,} &
    \multicolumn{2}{|c|}{Fixed points} &
    \makecell{Number of\\ relevant/ \\ irrelevant \\ operators}  & 
    \makecell{Anomalous dimensions} &
    \makecell{Coset space} & \makecell{IR stability}
    \\ \cline{1-2}
    \hline
    I &
     \makecell[l]{$\theta_c = \cos^{-1} \left(\pm \dfrac{1}{\sqrt{3}}  \right)$ \\[0.5em] $
      \phi_c = (2n+1)\dfrac{\pi}{4}$\\ $n=0,1,2,3$ } &

     \makecell[l]{$g^2_c = \epsilon - \dfrac{2}{3} \dfrac{\epsilon}{N_c} + \dfrac{\epsilon^2}{N_c} $ \\[0.5em] $ \lambda_c = \dfrac{4 \epsilon}{3 N_c}$}
     
     &
    \makecell{$n_{IRR} = 3, $\\$n_{R} = 0$}    & 
     
     \makecell[l]{$\eta_{\phi} = \epsilon - \dfrac{2}{3} \dfrac{\epsilon}{N_c} + \dfrac{1}{2} \dfrac{\epsilon^2}{N_c}$ \\[0.7em]$
    \eta_{\psi_i}= \dfrac{\epsilon}{3 N_c} - \dfrac{\epsilon^2}{4 N_c}$, \\[0.7em]$ i\in\{1,2,3\}$}
       &  
       \makecell{$(Z_2)^3 $}  &  

       IR stable
       
          \\  \hline

    II &
     \makecell[l]{ $(\theta_c,\phi_c) = \left( (2n+1)\dfrac{\pi}{4},m\dfrac{\pi}{2} \right)$\\[0.7em] $(\theta_c,\phi_c) = \left( \dfrac{\pi}{2}, (2n+1)\dfrac{\pi}{4} \right)$,\\[0.7em] $ m,n = 0,1,2,3$ } &

     \makecell[l]{$g^2_c = \epsilon - \dfrac{\epsilon}{N_c} + \dfrac{3}{2} \dfrac{\epsilon^2}{N_c}$ \\[0.7em] $\lambda_c = \dfrac{2\epsilon}{N_c} $}
     
      &  \makecell{$n_{IRR} = 2,$\\$n_{R} = 1$ }    &   
      \makecell[l]{ $\eta_{\phi} = \epsilon - \dfrac{\epsilon}{N_c} + \dfrac{3}{4} \dfrac{\epsilon^2}{N_c}$ \\[0.7em] $\eta_{\psi_i} = \eta_{\psi_{j}} =  \dfrac{\epsilon}{2N_c} - \dfrac{3}{8}\dfrac{\epsilon^2}{N_c}$  \\[0.7em]  $\eta_{\psi_{k\neq (i,j)}} =0$}   
                 &
        \makecell{$C_3 \otimes (Z_2)^2$} 

        &   IR unstable 
        
        \\ \cline{1-6} 

       III &
 
        \makecell{$\theta_c = 0,\pi$ \\ $(\theta_c. \phi_c) = \left(\pi/2, n\pi/2 \right)$ \\[0.7em] $n = 0,1,2,3$} &

        \makecell[l]{$g^2_c = \epsilon - \dfrac{2 \epsilon}{N_c} + \dfrac{3\epsilon^2}{N_c}$ \\[0.7em]
        $\lambda_c = \dfrac{4\epsilon}{N_c} $} &
        
             \makecell{$n_{IRR} = 1,$\\ $n_{R} = 2$} &
             
        \makecell{$ \eta_{\phi} = \epsilon - \dfrac{2 \epsilon}{N_c} + \dfrac{3}{2} \dfrac{\epsilon^2}{N_c}$ \\[0.7em] $\eta_{\psi_i} = \dfrac{\epsilon}{N_c} - \dfrac{3}{4} \dfrac{\epsilon^2}{N_c} $ \\[0.7em] $\eta_{\psi_{j}} = \eta_{\psi_{k}} = 0,$\\[0.7em]$\,  (i,j,k)\in \{1,2,3 \} $} &

        \makecell{$C_3 \otimes Z_2$} & 
        
        \\ \hline \hline
    \end{tabular}
    
\end{table*}

\begin{figure}[b]
\includegraphics[width=8.6cm, height=6.6cm]{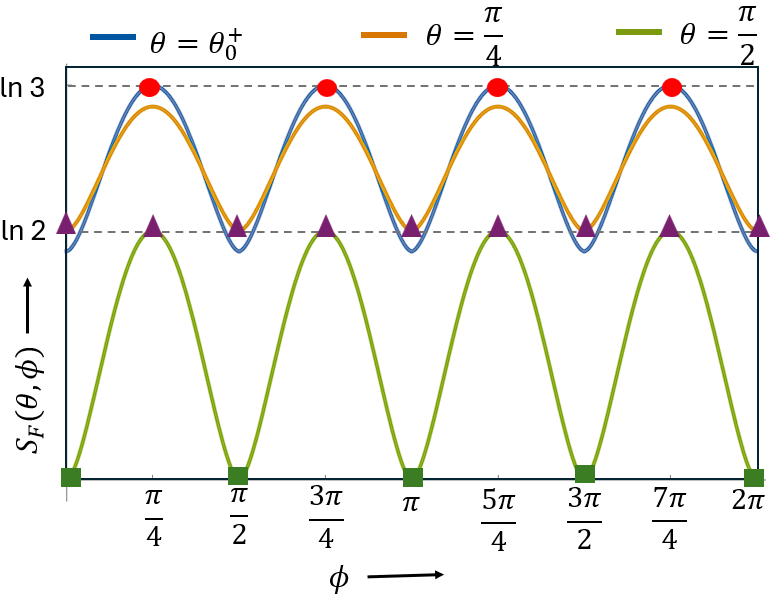}
\caption{Entanglement entropy, $S_{F}(\theta,\phi)$ (Eq. \ref{EentrpyN3}) of the fixed point interaction operator, $\mathcal{O}_{\text{Y}}$(Eq. (\ref{RdloprtrN3})) evaluated along the $S^2$ manifold for $\mathcal{N}=3$ flavor. It is plotted as a function of the azimuthal angle $\phi$  in the limit of $N_c \rightarrow \infty$. Here $S_F$ is plotted for three distinct values of the polar angles $\theta = \theta^{+}_0, \pi/4, \pi/2$ where $\theta^+_0 = \cos^{-1}1/\sqrt{3}$. The red circles denote the IR stable fixed points (\textit{class I}, Eq. (\ref{IRstbleN3})) that appear at first order in $1/N_c$. The pink triangles (\textit{class II}, Eq. (\ref{IRunstble1N3})) and green squares (\textit{class III}, Eq. (\ref{IRunstble2N3})) denote different classes of IR unstable fixed points appearing in the $O(1/N_c)$-expansion.}
\label{EentrpyN3}
\end{figure}

\subsection{$S^2$ Conformal manifold in the $N_c\rightarrow \infty$ limit}
 The flow along the polar and azimuthal directions, described by Eqs. (\ref{YkwaRGENcfnte3N2},\ref{YkwaRGENcfnte3N3}), vanishes for all $(\theta, \phi)$ in the $N_c \rightarrow \infty$ limit. Thus, a 2-sphere conformal manifold appears in the 3-dimensional flavor space described by the equation (Eq. (\ref{fxdpntsYkwalrgeNc})),
 \begin{eqnarray}
     g_c = \sqrt{\epsilon},
 \end{eqnarray}
same as the radius of the ring manifold in the $\mathcal{N}=2$ case. 
The conformal manifold has two exact marginal deformation operators at every point on the manifold. At a point, $(g_{1,c}, g_{2,c}, g_{3,c}) = (\sqrt{\epsilon} \sin \theta \cos \phi, \sqrt{\epsilon} \sin \theta \sin \phi, \sqrt{\epsilon} \cos \theta )$ on the manifold, the exact marginal operators read,
\begin{subequations}
\begin{eqnarray}
    \mathcal{O}_{\text{D}1}(\theta, \phi) &=& \phi^*\bigl( \cos \theta \cos \phi \,\, \psi^{T}_{1}\left(i s_y \right) \psi_{1} + \nonumber \\ && \cos \theta \sin \phi \,  \psi^{T}_{2}\left(i s_y \right) \psi_{2} - \sin \theta  \,\ \psi^{T}_{3}\left(i s_y \right) \psi_{3}  \bigr)\nonumber \\ &+& \text{h.c} \label{Mgnloprtr1N3} \\
    \mathcal{O}_{\text{D}2}(\theta, \phi) &=& - \sin \phi \,\, \psi^{T}_{1}\left(i s_y \right) \psi_{1} \phi^{*} \nonumber \\ &+& \cos \phi \,\, \psi^{T}_{2}\left(i s_y \right) \psi_{2} \phi^{*} + \text{h.c} \label{Mgnloprtr2N3}\\
    \Delta_{\mathcal{O}_{\text{D}1}} &=& \Delta_{\mathcal{O}_{\text{D}2}} = d+1 \nonumber
\end{eqnarray}    
\end{subequations}
These two exactly marginal operators define a 2-dimensional tangent space at the point $(\theta, \phi)$ on the manifold. Note that any operator that lies on this tangent space is also exactly marginal. 

The fixed point interaction operator of the 2-sphere manifold is given by,
\begin{eqnarray}
     \mathcal{O}_{\text{Y}}(\theta, \phi) &=& \phi^*\bigl( \sin \theta \cos \phi \,\, \psi^{T}_{1}\left(i s_y \right) \psi_{1} + \nonumber \\ && \sin \theta \sin \phi  \,\psi^{T}_{2}\left(i s_y \right) \psi_{2} + \cos \theta  \,\ \psi^{T}_{3}\left(i s_y \right) \psi_{3}  \bigr)\nonumber \\ &+& \text{h.c} \label{RdloprtrN3}  \\
     \Delta_{\mathcal{O}_{\text{Y}}} &=& 4. \nonumber
\end{eqnarray}

Similar to the $\mathcal{N}=2$ flavor case, both the marginal and interaction operators are functions of the manifold coordinates $(\theta, \phi)$. We can also observe an EPR-like entanglement structure in these operators in the flavor space, which varies with the manifold coordinates.

We calculate the entanglement entropy $S_F$ of the fixed point interaction operator $\mathcal{O}_{\text{Y}}$, using the formula in Eq. (\ref{Eentrpy}). We have,
\begin{eqnarray}
    S_{F}(\theta, \phi) &=& -\bigl[ \sin^2\theta \cos^2\phi \ln \left(\sin^2\theta \cos^2\phi\right) \nonumber \\ &+&  \sin^2\theta \sin^2\phi \ln \left(\sin^2\theta \sin^2\phi \right) + \cos^2\theta \ln \cos^2\theta \bigr]\nonumber \\ \label{EentrpyN3}
\end{eqnarray}
 In Fig. \ref{EentrpyN3}, we plot $S_{F}(\theta, \phi)$ as a function of the coordinates $(\theta, \phi)$. 
 
 We notice that the entanglement entropy of the operator is maximum ($=\ln 3$) at points 
\begin{equation}
(\theta^{\pm}_0, \phi_{2k-1}) = \left(\cos^{-1}\left(\pm \frac{1}{\sqrt{3}}\right),(2k-1)\frac{\pi}{4} \right),   
\end{equation}
for $k=1,2,3,4$ on the manifold. As an example, the interaction operators at the point $(\theta^{+}_0,\phi_1 )$ has the form,
\begin{eqnarray}
   \mathcal{O}_{\text{Y}}\left(\theta^{+}_0,\phi_1  \right) &=&  \frac{1}{\sqrt{3}} \phi^{*} \bigl( \psi^{T}_{1}\left(i s_y \right) \psi_{1} + \psi^{T}_{2}\left(i s_y \right) \psi_{2} \nonumber \\ &+& \psi^{T}_{3}\left(i s_y \right) \psi_{3} \bigr) + \text{h.c}
   \\
   S_{F}(\theta^{+}_0,\phi_1) &=& \ln 3, \nonumber
\end{eqnarray}
maximally entangled in the flavor space.

However, at the points 
\begin{equation}
(\theta, \phi_{2k}) = (\pi/2, k\pi/2)\,\, \text{or} \,\, \theta = 0,\pi,
\end{equation}
The entanglement entropy of the interaction operator is zero. For example, at $(\theta,\phi) = (0,0)$, 

\begin{eqnarray}
    \mathcal{O}_{\text{Y}}(0,0) &=& \psi^{T}_{3}\left(i s_y \right) \psi_{3} \phi^{*} + \text{h.c}. \label{N3lrgeNcprdct} \\
    S_{F}(0,0) &=&  0. \nonumber
\end{eqnarray}    
The operator is in a product state in the flavor space.

As in the $\mathcal{N}=2$ case, the entanglement structure of the operators indeed is a function of the manifold coordinates. We also observe that the maxima of the entanglement entropy scale as $S_{F}\sim \ln \mathcal{N}$, for $\mathcal{N}=2,3$.

\subsection{Isolated Wilson-Fisher fixed points at $O(1/N_c)$}

The $O(1/N_c)$ corrections induce RG flow along the azimuthal and polar direction on the 2-sphere manifold, as Eqs. (\ref{YkwaRGENcfnte3N2},\ref{YkwaRGENcfnte3N3}) show, driven by the fermion field renormalization effect. As a result, the manifold breaks into $3^3 - 1 = 26$ isolated strong-coupling fixed points, which can be categorized into three distinct universality classes, namely,
\begin{itemize}
    \item  Class I: 
    \begin{equation}
        (\theta_c, \phi_c) = \left(\cos^{-1} \left(\pm \dfrac{1}{\sqrt{3}}  \right), (2n+1)\dfrac{\pi}{4}\right) \label{IRstbleN3}
    \end{equation},  
    \item  Class II: 
    \begin{equation}
     (\theta_c,\phi_c) = \left( (2n+1)\dfrac{\pi}{4},m\dfrac{\pi}{2} \right)   \label{IRunstble1N3}
    \end{equation} 
    \item Class III:
    \begin{eqnarray}
     \theta_c = 0,\pi\,\, \text{and} \,\, (\theta_c. \phi_c) = \left(\pi/2, n\pi/2 \right)   \label{IRunstble2N3}
    \end{eqnarray}
\end{itemize}
where $n,m \in \{0,1,2,3 \}$.
We discuss them one by one below,

\paragraph*{ \textbf{Class I}:} In this universality class, all three fermion flavors are strongly coupled to the bosons at an equal magnitude. Expanding the critical Yukawa coupling strengths in Cartesian form to order $O(\epsilon^2, 1/N_c)$,

\begin{equation}
  g^2_{i,c} = \frac{\epsilon}{3}  - \dfrac{2}{9} \dfrac{\epsilon}{N_c} + \dfrac{\epsilon^2}{3N_c} \,\,\text{for} \,\, i=1,2,3.  
\end{equation}
It is evident that all three couplings, $g_{i,c}$, are equal in magnitude. 

$H = S_3$ is the invariant subgroup of the set of fixed points. Here $S_3$ stands for the permutation group of size $3$.
From Eq. (\ref{smmtrygrpRGEgen}), we find that the symmetry group of the RGE in Eq. (\ref{YkwaRGENcfntegen1}) for $\mathcal{N}=3$ is $G = S_3 \otimes (Z_2)^3$. Therefore, the set of fixed points forms elements of the coset space, 
$$C = \frac{S_3 \otimes (Z_2)^3}{S_3} =(Z_2)^3$$
with $N(C) = 8$ isolated fixed points. The table \ref {ykwaN3tble} lists the fixed points $ g^2_c$ and $ \lambda_c$, and the anomalous dimensions, expanded to the order $O(\epsilon^2, 1/N_c)$. 

This set of fixed points defined in Eq. (\ref{IRstbleN3}) turns out to be infrared stable. The tangential deformation operators defined by Eqs. (\ref{Mgnloprtr1N3},\ref{Mgnloprtr2N3}) at these points (Eq. (\ref{IRstbleN3})), which were exactly marginal in the $N_c \rightarrow \infty$ limit, now become weakly irrelevant at finite $N_c$. 

We observe that the interaction operator, defined in Eq. (\ref{RdloprtrN3}), at these fixed points (Eq. (\ref{IRstbleN3})) is maximally entangled in the flavor space (red circles in Fig. \ref{EentrpyN3}). 
 Table \ref{YkwaN3dfmtn} shows the operator along with its scaling dimensions and its entanglement entropy at a selected point in this universality class. 

Similar to the $\mathcal{N}=2$ flavor case, we see that the finite $N_c$ renormalization group flows are to those points where the fixed point interaction operators are maximally entangled in the flavor space, therefore dictating the surface topological quantum criticality(sTQCP).

\begin{table}[H]
 \caption{Yukawa interaction operator at selected points in the three universality classes of an $\mathcal{N}=3$ flavor TI surface, when studied at first order in $1/N_c$. We list their scaling dimensions, expanded to $O(\epsilon^2, 1/N_c)$ and the entanglement entropy, $S_F$ (Eq. (\ref{EentrpyN2})). Here $\theta^+_0 = \cos^{-1}1/\sqrt{3}$.} 
 \label{YkwaN3dfmtn} 
   \renewcommand{\arraystretch}{4.0}
\begin{tabular}{|c|c|c|c|c|}
\hline \hline
\makecell{\,}  &
\makecell{ Fixed \\ point }& \makecell{Yukawa \\ interaction operator}  & \makecell{Scaling \\ dimens-\\ions} & \makecell{Entang-\\lement\\ entropy\\ ($S_F$)}
 \\ \hline
 I&
 \makecell{$\theta_c = \theta^+_0 $\\[1.0em]$ \phi_c = \dfrac{\pi}{4} $} & 
 
 \makecell{$ \dfrac{\phi^{*}}{\sqrt{3}} [\psi^{T}_{1}\left(i s_y \right) \psi_{1} +$ \\[0.5em] $ \psi^{T}_{2}\left(i s_y \right) \psi_{2} +\psi^{T}_{3}\left(i s_y \right) \psi_{3} \bigr] $\\[0.5em]  + h.c }

 & \makecell{$4 - \dfrac{\epsilon^2}{N_c} $}  & $\ln 3$ 
 
 \\    \hline

 II  & 

\makecell{$\theta_c = \dfrac{\pi}{2}$ \\[0.7em] $\phi_c = \dfrac{\pi}{4}$}  &

   \makecell{$ \dfrac{\phi^{*}}{\sqrt{2}} [ \psi^{T}_{1}\left(i s_y \right) \psi_{1} +$\\[1.0em]$  \psi^{T}_{2}\left(i s_y \right) \psi_{2}]$ + h.c}  &

   \makecell{$4  -  \dfrac{3\epsilon^2}{2 N_c}$} & 
   
   $\ln 2$ \\
       \hline

   III  &
   
   \makecell{$\theta_c = 0$ \\ $\phi_c = 0$} &

    \makecell{$\phi^{*} \psi^{T}_{3}\left(i s_y \right) \psi_{3}$+ h.c}  &

    \makecell{$4 - \dfrac{3\,\epsilon^2}{N_c} $}  &

    $0$\\ \hline \hline
 \end{tabular}
\end{table}

\paragraph*{ \textbf{Class II}:} Only two out of three fermion flavors are coupled to bosons in this universality class; the third one is free. In terms of the Cartesian coordinates,
\begin{eqnarray}
  g^2_{i,c} = g^2_{j,c} = \frac{\epsilon}{2} - \dfrac{\epsilon}{2N_c} + \dfrac{3}{4} \dfrac{\epsilon^2}{N_c},\,\,\, g^2_{k,c} =0 
\end{eqnarray}
 for distinct  $(i,j,k) \in \{1,2,3\}$(see Table (\ref{ykwaN3tble}) for detailed solutions). 
 
 As before, we find the invariant subgroup $H = S_{2} \otimes Z_2$. Hence, the coset space is $$C = C_3 \otimes (Z_2)^2,$$ where $C_3$ stands for cyclic group of size $3$. We have $N(C) = 12$ elements in this universality class. 

 At these fixed points (Eq. (\ref{IRunstble1N3})), while one of the deformation operators (Eqs. \ref{Mgnloprtr1N3},\ref{Mgnloprtr2N3}) has become weakly irrelevant, the other turned out to be weakly relevant in the flavor space.

 The entanglement entropy of the interaction operator (\ref{RdloprtrN3}) at these fixed points (Eq. (\ref{IRunstble1N3})) is $S_{F} = \ln 2$ (pink triangles in Fig. \ref{EentrpyN3}, also listed in Table.(\ref{YkwaN3dfmtn}). Although the operator remains entangled, this value is smaller than the entanglement entropy at the IR stable fixed points discussed previously (class I).

\paragraph*{ \textbf{Class III}:} In this universality class, only one fermion flavor is coupled to the bosons, the rest two are free. That is,
\begin{equation}
  g^2_{i,c} = \epsilon - \dfrac{2 \epsilon}{N_c} + \dfrac{3}{2} \dfrac{\epsilon^2}{N_c}, \,\, 
g^2_{j,c}=g^2_{k,c} = 0,  
\end{equation}

 for distinct $(i,j,k) \in \{1,2,3 \}$(see Table \ref{ykwaN3tble}). We find the invariant subgroup $H = S_2 \otimes (Z_2)^2$ and hence the coset space $$C = C_3 \otimes Z_2$$. Therefore, we have $N(C) = 6$ elements in this universality class.

Both the deformation operators defined in Eqs.  (\ref{Mgnloprtr1N3}, \ref{Mgnloprtr2N3}) at these points (identified by Eq.(\ref{IRunstble2N3})) become weakly relevant in this case. Therefore, the universality class is infrared unstable.

Also, we find that the entanglement entropy of the fixed point interaction operator is zero. In other words, the operator is essentially in a product state in the flavor space (green squares in Fig. \ref{EentrpyN3}, also listed in table \ref{YkwaN3dfmtn}).

Also, note that this set of IR unstable fixed points is equivalent to the SUSY fixed points in the $\mathcal{N}=1$ case. However, similar to the $\mathcal{N}=2$ case discussed previously, and the fixed points do not define the physics of $\mathcal{N}=3$ sTQCP.

The conclusive remarks are similar to the $\mathcal{N}=2$ case studied in Sec. \ref{sec:N2Ykwa}. The $S^2$ conformal manifold observed at $N_c\rightarrow \infty$ limit breaks down into $26$ isolated fixed points, grouped into three distinct classes, once finite-$N_c$ quantum fluctuations are included. Of these, the universality class of the surface topological quantum criticality(sTQCP) is characterized by the maximally entangled Yukawa interaction operators in the flavor space, with all three Dirac cones strongly coupled to the bosons.

\section{Conclusion}

In conclusion, we have studied surface topological quantum critical points (sTQCP) with $\mathcal{N}=2,3$ surface Dirac cones.
In our studies, we have employed bulk topologies to determine the degrees of freedom at the surface TQCP and applied the protecting symmetry $G_p=Z^T_2$ to define the structure of interaction matrices or parameters. 
Although our studies are motivated by surface TQCP and are mainly presented in this context, the idea and approach we applied to derive our conclusions apply to other general studies of sTQCPs where the dimension of the interaction parameter space $D_p$ is larger or much larger than one, i.e., a high-dimensional parameter space.

The main general conclusions are:

I) Identifying a conformal manifold, i.e., a smooth manifold of scale-conformal invariant theories in some limit, can be a very powerful approach if the interaction parameter space of sTQCPs is higher or much higher than one. In our studies, the conformal manifold appears naturally in the limit where there is an infinite number of colors of surface fermions, i.e., $N_c \rightarrow \infty$.

A phase boundary generically is a co-dimension one manifold in a parameter space of general dimension $D_p$ (where $D_p=\mathcal{N}+1$, $\mathcal{N} =2,3,...$ in our studies), i.e., a manifold of dimension $D_p-1$.

On the other hand, we have illustrated that the conformal manifold is usually a sub-manifold further embedded in the phase boundary manifold, as not all the theories appearing in the (mean-field) phase boundary
are fully scale invariant. In the class of sTQCP models we have studied, a conformal manifold is always a co-dimension two sub-manifold in the parameter space, i.e., a manifold with dimension $D_p-2$. So a conformal manifold is one dimension lower than the phase boundary manifold.

II) In generic sTQCPs, the conformal manifolds emerging in certain limits are unstable when subject to various higher-order quantum fluctuations. In our studies, instability does appear when $1/N_c$ quantum corrections are adequately taken into account.

These quantum effects typically further induce weak flows within the conformal manifolds obtained in the infinite $N_c$-limit. Smooth manifolds of dimension $\mathcal{N}-1$ (where $\mathcal{N}$ is the number of surface Dirac cones) break down into an exponentially large number of fixed points. For $\mathcal{N}=2$, the number of fixed points that emerge in the conformal manifold is 8, with four of them IR stable and the rest being IR unstable.
For $\mathcal{N}=3$, the number becomes 26, with eight IR-stable fixed points.

 In general, there will be a total $3^{\mathcal{N}}-1$ fixed points, among which $2^{\mathcal{N}}$ fixed points can be shown to be IR stable.
 
 III) 
Only the IR stable fixed points define the universality of sTQCPs. We are able to connect the stability of these fixed points with a flavor-space entanglement entropy $S_F$. The IR stable fixed points (along all directions tangent to the manifold ) have maximal entanglement entropy, while the stability decreases as the entropy is lowered.

The fixed points of zero entanglement entropy effectively only involve one single surface Dirac cone and are IR unstable along all directions within the manifold. So they also exhibit the well-known emergent supersymmetry. However, they
do not play a role at sTQCPs under our considerations in this article, where the interaction parameter space is always higher than one and there is more than one Dirac cone. 

A few open questions we plan to pursue in the future. The first one is to generalize our current result to an arbitrary number of flavors $\mathcal{N}$ and present the details of RGE flows.
And to examine explicitly the correlations between the direction of the flows and entanglement entropy. This is currently under pursuit by the authors, and all the observations made in this article appear to be valid in more general cases. A more thorough examination will be presented soon.

The second question is the relation between our studies here and the general $F$-theorem in $2+1$D CFT, where it was shown that along the RGE flows the $S^3$ CFT free energy $Z_{S^3}$ shall always 
decrease\cite{Jafferis(2011),klebanov(2011),jafferis(2012)}. The $F$-theorem statement, along with its intimate connection to a constant term (but sub-leading) in the Von Neumann entropy of a CFT\cite{Casini(2011),sachdev(2017)}, has served as an analogue of Zamaldochikov's c-theorem introduced in $(1+1)D$ CFT.
We plan to further explore the possible connections between sTQCPs and the $F$-theorem in the future to better understand the utility of theories of entanglement entropy in our studies of topological gapless states.

F.Z. wants to thank Cenke Xu and Hirosi Ooguri for inspiring discussions on classifications of conformal operators and manifolds, and entanglement entropy. This project is in part supported by an NSERC (Canada) discovery grant under grant number RGPIN-2020-07070.

\appendix

\section{Renormalization group analysis\label{appndx:RGE}}

Here, we present the derivation of the renormalization group equation in Eqs. (\ref{YkwaRGENcfntegen1},\ref{YkwaRGENcfntegen2}). We use the minimal subtraction scheme to derive the RGE\cite{justinbook(2002),sachdev(2011),Shankar(2017),fradkinbook(2019)}. First, we identify the action shown in Eq. (\ref{YkwEFTNc}) as the bare action with bare fields $\psi_{i,\alpha,0}$, $\phi_0$ and couplings $g_{i,0}$ and $\lambda_0$. We write down the renormalized Lagrangian below as,
\begin{eqnarray}
    \mathcal{L}_{R} &=&  \sum_{i=1}^{\mathcal{N}}\sum_{\alpha=1}^{N_c} Z_{\psi_i} \psi^{\dagger}_{i,\alpha,R} \left[ \partial_\tau + i s_y \partial_x - i s_x\partial_y \right]\psi_{i,\alpha,R}\nonumber \\ &+&  Z_{\phi}|\partial_{\tau} \phi_R |^2 + Z_{\phi}\sum_{i=1}^{d}|\partial_{i} \phi_R |^2 + Z^2_{\lambda}\tilde{\lambda}_R \mu^{\epsilon} |\phi_R|^4 \nonumber \\  &+&
    \sum^{\mathcal{N}}_{i = 1} \sum_{\alpha=1}^{N_c}  \frac{Z_g \tilde{g}_{i,R}}{\sqrt{N_c}}\mu^{\epsilon/2} \phi^{*}\,\psi^{T}_{i,\alpha,R} i s_y \psi_{i,\alpha,R} +\text{h.c.}
\end{eqnarray}
Here, the bare and the renormalized fields and couplings are related by the following set of equations,
\begin{eqnarray}
    \psi_{i,\alpha,0} &=& \sqrt{Z_{\psi_{i}}}\psi_{i,\alpha,R}, \,\, \phi_0 = \sqrt{Z_{\phi}}\, \phi_R \label{RGcndtns} \\
    Z_{g}\tilde{g}_{i,R} &=& \mu^{-\epsilon/2} g_{i,0} Z_{\psi_i} \sqrt{Z_{\phi}} \nonumber \\
    Z_{\lambda}\tilde{\lambda}_R &=& \mu^{-\epsilon} \lambda_0 Z^2_{\phi}  \nonumber
\end{eqnarray}

We showed in Fig. \ref{vrtxrnmlztn} that the Yukawa vertex renormalization starts appearing only at order $O(1/N^2_c)$. That is, $$Z_g = 1 + O(1/N^2_c).$$Therefore, the beta function for the Yukawa coupling simplifies to,
\begin{eqnarray}
     \frac{d\tilde{g}_{i,R}}{d\ln\mu} &=& -\frac{\epsilon}{2} \tilde{g}_{i,R} + \tilde{g}_{i,R} \left( \eta_{\psi_i} + \frac{1}{2} \eta_{\phi} \right) \label{appndxbtraykwa}
\end{eqnarray}
where we used the fact that the bare coupling $g_{i,0}$ is independent of the scale $\mu$.  $\eta_{\phi}$ and $\eta_{\psi_i}$  are the anomalous dimensions of the boson and fermion fields, respectively, with the following definition,
 \begin{eqnarray}
    \eta_{\alpha} &=& \frac{\partial 
    \ln Z_{\alpha}}{\partial \ln \mu}\label{appndxanmlsdmnsn} \\ &=& \sum_{j} \beta_{\tilde{g}_{j,R}} \frac{\partial \ln Z_{\alpha} }{\partial \tilde{g}_{j,R}} + \beta_{\tilde{\lambda}_{R}} \frac{\partial \ln Z_{\alpha} }{\partial \tilde{\lambda}_{R}} , \,\,\,   \alpha = \phi, \psi_i.\nonumber
\end{eqnarray}
In the last step, we recast the partial derivatives in terms of the beta functions of the couplings, which are more suited for the minimal subtraction scheme we use to derive the RG equations.  

\subsection{Field renormalization}
We study the boson and fermion field renormalization up to two-loop level and up to first order in $1/N_c$. We write down the bare two-point function for the boson and fermion fields below,
\begin{eqnarray}
    \Gamma^{(2)}_{\phi,0} &=& \textbf{k}^2 + \Pi^{(1)}(\textbf{k}) + \Pi^{(2)}(\textbf{k})\label{appndxbare2pntbsn} \\
    \Gamma^{(2)}_{\psi_i,0} &=& -ik_0 + \vec{s}.\vec{k} + \Sigma^{(1)}_{i}(\textbf{k}) + \Sigma^{(2)}_{i}(\textbf{k}) \label{appndxbare2pntfrmn}
\end{eqnarray}
Here, $\Pi^{(n)}(\textbf{k})$ ($\Sigma^{(n)}_{i}(\textbf{k})$) represent the boson (fermion) self-energy at $n$th loop order, where $n=1,2$.

We first look at the one-loop corrections to both the fields. 
The boson self-energy  $\Pi^{(1)}(\textbf{k})$ is due to the fermion bubble in Fig. \ref{dgrmslrgeNc}(a), which is finite at $N_c \rightarrow \infty$. The fermion field correction $\Sigma^{(1)}_{i}(\textbf{k})$, shown in Fig. \ref{dgrmslrgeNc}(b), start appearing only at first order in $1/N_c$. Below, we write down their integral expressions in terms of the fermion and boson Green's functions. Then, we expand the integrals in powers of $\epsilon = 4 - D$, where $D = 1 +d$ is the spacetime dimensions.
\begin{eqnarray}
    \Pi^{(1)}(\textbf{k}) &=& -2 \sum^{\mathcal{N}}_{i=1} g^2_{i,0} \int \frac{d^D q}{(2\pi)^D} \text{Tr} \left[ G_i (\textbf{q}) s_y G^T_i (\textbf{k} - \textbf{q}) s_y \right] \nonumber \\ &=& \textbf{k}^2\sum^{\mathcal{N}}_{i=1} g^2_{i,0} \,  \left(\frac{1}{4\pi^2 \epsilon} + c_{\phi,1} \right)  + O(\epsilon) \\
    \Sigma^{(1)}_i(\textbf{k}) &=& \frac{4g^2_{i,0}}{N_c} \int \frac{d^Dq}{\left(2\pi \right)^D} D(\textbf{q})s_y G_i^T(\textbf{q} - \textbf{k})s_y \nonumber \\ &=& \frac{g^2_i}{N_c} (-ik_0 +\vec{s}.\vec{k})\biggl(\frac{1}{4\pi^2 \epsilon} + c_{\psi,1} + O(\epsilon)\biggr) 
    \end{eqnarray}
where $c_{\phi,1}$ and $c_{\psi,1}$ are numerical constants given by,
\begin{eqnarray}
    c_{\phi,1} &=& c_{\psi,1} = \frac{2 - \gamma + \ln(4\pi)}{8\pi^2} \nonumber 
\end{eqnarray}
Here, $G_{i}(\textbf{k})$ is the fermion Green's function for flavor $i$, while $D(\textbf{k})$ is the boson Green's function. 

At the two-loop level, the sunrise diagram in Fig. \ref{dgrmslrgeNc}(e) contributes to the boson self-energy $\Pi^{(2)}(\textbf{k})$, while the diagram in Fig. \ref{dgrmslrgeNc}(d) contributes to the fermion self-energy $\Sigma_i^{(2)}(\textbf{k})$. They are given by,
    \begin{eqnarray}
    \Pi^{(2)}(\textbf{k}) &=& -4 \sum^{\mathcal{N}}_{i=1} g^2_{i,0} \int \frac{d^D \textbf{q}}{(2\pi)^D} \\  &&\text{Tr} \left[ G_i (\textbf{q}) \Sigma^{(1)}_i (\textbf{q}) G_i (\textbf{q}) s_y G^T_i (\textbf{k} - \textbf{q}) s_y \right] \nonumber \\ &=& -\textbf{k}^2 \sum^{\mathcal{N}}_{i=1} \frac{g^2_{i,0}}{N_c} \left( \frac{1}{16 \pi^4 \epsilon^2} + \frac{c_{\phi,2}}{\epsilon} \right) + O(\epsilon^0)   \nonumber   
    \end{eqnarray}    
    \begin{eqnarray}
    \Sigma^{(2)}_{i}(\textbf{k}) &=&  \frac{4g^2_{i,0}}{N_c} \int \frac{d^D \textbf{q}}{\left(2\pi \right)^D} \\  && D (\textbf{q}) \Pi^{(1)} (\textbf{q}) D (\textbf{q}) s_y G_i^T(\textbf{q} - \textbf{k})s_y \nonumber \\  &=& - (-ik_0 + \vec{s}.\vec{k}) \frac{g^4_{i,0}}{N_c} \sum^{\mathcal{N}}_{j=1} g^2_{j,0} \left( \frac{1}{32 \pi^4 \epsilon^2} + \frac{c_{\psi,2}}{\epsilon} \right)\nonumber
\end{eqnarray}
where $c_{\phi,2}$ and $c_{\psi,2}$ are numerical constants given by,
\begin{eqnarray}
    c_{\phi,2} &=& 2c_{\psi,2}= \frac{11-4\gamma + 4 \ln (4 \pi) }{64\pi^4}. \nonumber 
\end{eqnarray}

Plugging these back into Eqs. (\ref{appndxbare2pntbsn},\ref{appndxbare2pntfrmn}), we get the bare 2-point functions in terms of bare couplings. Now, we define the renormalized two-point functions as, 
\begin{eqnarray}
    \Gamma^{(2)}_{\phi,R} &=& Z_{\phi} \Gamma^{(2)}_{\phi,0} \nonumber \\ 
     \Gamma^{(2)}_{\psi_i, R} &=& Z_{\psi_i} \Gamma^{(2)}_{\psi_i, 0} 
\end{eqnarray}
$\Gamma^{(2)}_{\phi,R}$ and $\Gamma^{(2)}_{\psi_i,R}$ in terms of the renormalized couplings becomes,
\begin{widetext}
    \begin{eqnarray}
        \Gamma^{(2)}_{\phi,R} &=& \textbf{k}^2 \left[Z_{\phi} +   \left(\frac{1}{4\pi^2 \epsilon} + c_{\phi,1} \right) \sum^{\mathcal{N}}_{i=1} \tilde{g}^2_{i,R}\mu^{\epsilon}Z^{-2}_{\psi_i}  - \left( \frac{1}{16\pi^4 \epsilon^2} + \frac{c_{\phi,2}}{\epsilon} \right)\sum_{i} \frac{\tilde{g}^4_{i,R}\mu^{2\epsilon}Z^{-1}_{\phi}Z^{-4}_{\psi_i}}{N_c} \right] \\
        \Gamma^{(2)}_{\psi_i,R} &=& \left(  -ik_0 + \vec{s}.\vec{k} \right) \left[Z_{\psi_i} + \left(\frac{1}{4\pi^2 \epsilon} + c_{\psi,1} \right) \frac{\tilde{g}^2_{i,R}\mu^{\epsilon} \left(Z_{\phi}Z_{\psi_i}\right)^{-1} }{N_c} - \left(\frac{1}{32\pi^4\epsilon^2} + \frac{c_{\psi,2}}{\epsilon}  \right) \frac{\tilde{g}^2_{i,R}}{N_c} \sum_{j} \tilde{g}^{2}_{j,R} Z^{-3}_{\psi_i}Z^{-2}_{\phi} \right]
    \end{eqnarray}
\end{widetext}
We want the renormalized two-point functions to be finite up to the two-loop level and first order in $1/N_c$. So the field renormalization factors $Z_{\phi}$ and $Z_{\psi_i}$ must absorb all the poles in $\epsilon$, both first order and second order. We find that both $\Gamma^{(2)}_{\phi,R}$ and $\Gamma^{(2)}_{\psi_i,R}$ become finite and smooth functions of couplings when,
\begin{eqnarray}
    Z_{\phi} &=& 1 -\frac{1}{4\pi^2 \epsilon} \sum_{i} \tilde{g}^2_{i,R} +  \left(\frac{3}{64 \pi^4\epsilon} - \frac{1}{16\pi^4\epsilon^2} \right) \sum_{i} \frac{\tilde{g}^4_{i,R}}{N_c} \nonumber \\ \\
 Z_{\psi_i} &=& 1 - \frac{1}{4\pi^2\epsilon} \frac{\tilde{g}^2_{i,R}}{N_c} + \left( \frac{3}{128\pi^4\epsilon} - \frac{1}{32\pi^4\epsilon^2} \right) \frac{\tilde{g}^2_{i,R}}{N_c} \sum_{j} \tilde{g}^{2}_{j,R} \nonumber \\
\end{eqnarray}
up to the two-loop level and first order in $1/N_c$. Using the formula in Eq. (\ref{appndxanmlsdmnsn}), we find that the anomalous dimensions of the bosons and the fermions become,
\begin{eqnarray}
    \eta_{\phi} &=& \frac{1}{4\pi^2} \sum_i \tilde{g}^2_{i,R} - \frac{3}{32 \pi^4}\sum_i\frac{\tilde{g}^4_{i,R}}{N_c} \\
    \eta_{\psi_i} &=& \frac{1}{4\pi^2} \frac{\tilde{g}^2_{i,R}}{N_c} - \frac{3}{64\pi^4} \frac{\tilde{g}^2_{i,R}}{N_c}\sum_k\tilde{g}^2_{k,R}    
\end{eqnarray}
Notice that the anomalous dimensions of the fields are smooth functions of the coupling strengths, although the renormalization factors had poles in $\epsilon$. 
After rescaling $g^2_i/4\pi^2 \rightarrow g^2_i$, we arrive at the Eqs. (\ref{anmlsfnteNcbsn},\ref{anmlsfnteNcfrmn}) shown in the main text. 

Armed with the anomalous dimensions of the fields, the beta function of the Yukawa coupling can be derived in a straightforward manner using Eq. (\ref{appndxbtraykwa}). We get,

\begin{eqnarray}
    \frac{d\tilde{g}_{i,R}}{d\ln\mu} &=&    -\frac{\epsilon}{2} \tilde{g}_{i,R} + \frac{1}{8\pi^2} \tilde{g}_{i,R} \sum_j \tilde{g}^2_{j,R} + \frac{1}{4\pi^2} \frac{\tilde{g}^3_{i,R}}{N_c} \nonumber \\ &-& \frac{3}{64 \pi^4} \tilde{g}_{i,R} \sum_j \frac{\tilde{g}^4_{j,R}}{N_c} - \frac{3}{64\pi^4} \frac{\tilde{g}^3_{i,R}}{N_c}\sum_j\tilde{g}^2_{j,R}\nonumber
\end{eqnarray}
While writing the beta function in Eq. (\ref{YkwaRGENcfntegen1}) in the main text, we used $g_i$ to denote the dimensionless renormalized coupling instead of $\tilde{g}_{i,R}$ for brevity. We redefined the coupling as $g^2_i/4\pi^2 \rightarrow g^2_i$. Also, the flow parameter $\mu$ is redefined as $\mu = e^{-l}$ so that the flow in Eq. (\ref{YkwaRGENcfntegen1}) is towards the infrared limit.  

\subsection{Four-boson vertex renormalization}

The lowest order correction to the four-boson vertex appears at first order in $1/N_c$, as shown in Fig.\ref{dgrmslrgeNc}(c). The bare four-point correlation function has the form,
\begin{eqnarray}
    \Gamma^{(4)}_{0} = \lambda_{0} + \delta_g \lambda
\end{eqnarray}
where $\delta_g\lambda$ has the contributions from the diagram in Eq. (\ref{dgrmslrgeNc})(c). It has the form,
\begin{eqnarray}
    \delta_g \lambda &=& 4 \sum_{i} \frac{g^4_{i,0}}{N_c} \int \frac{d^Dq}{\left(2\pi \right)^D} 
    \nonumber\\ &&\text{Tr}\left[G_i \left(\textbf{q}\right) s_y G^T_i \left(\textbf{k} -\textbf{q}\right) s_y G_i \left(\textbf{q}\right) s_y G^T_i \left( -\textbf{k} -\textbf{q}\right)s_y \right]\nonumber \\ 
    &=& \frac{4}{4\pi^2 \epsilon} \sum_{i} \frac{g^4_{i,0}}{N_c} \frac{1}{\textbf{k}^{4-D}} + O(\epsilon^0) 
\end{eqnarray}
Here, we set the centre-of-mass momentum of incoming and outgoing particles to zero for brevity.

Using the relations in Eq. (\ref{RGcndtns}), the renormalized four-point function is given by,
\begin{eqnarray}
    \Gamma^{(4)}_{R} &=& Z_{\phi}^2 \Gamma^{(4)}_0
\end{eqnarray}
$\Gamma^{(4)}_{R}$ in terms of renormalized couplings become,
\begin{eqnarray}
     \Gamma^{(4)}_{R} &=& \tilde{\lambda}_R Z_{\lambda} + \frac{4}{4\pi^2\epsilon} \sum_i \frac{\tilde{g}^4_{i,R}}{N_c} + O(\epsilon^0)
\end{eqnarray}

Demanding the $\Gamma^{(4)}_{R}$ is finite, the vertex renormalization factor $Z_{\lambda}$ must be,
\begin{eqnarray}
     Z_{\lambda} = 1 - \frac{4}{4\pi^2 \epsilon} \frac{1}{\tilde{\lambda}_R} \sum_i \frac{\tilde{g}^4_{i,R}}{N_c} 
\end{eqnarray}
Using the relation between bare and renormalized couplings in Eq. (\ref{RGcndtns}) and using the fact that $\lambda_0$ is independent of scale $\mu$, we arrive at the following beta function for $\lambda$,
\begin{eqnarray}
     \frac{d \ \tilde{\lambda}_R}{d \ln \mu} &=& - \epsilon \tilde{\lambda}_R + \tilde{\lambda}_R \left( 2 \frac{d\ln Z_{\phi}}{d \ln \mu} - \frac{d \ln Z_{\lambda}}{d \ln \mu} \right) \\ 
     &=&   \tilde{\lambda}_R \left( - \epsilon  + \frac{1}{2\pi^2} \sum_j \tilde{g}^2_{i,R} \right) - \frac{4}{4\pi^2 } \sum_i \frac{\tilde{g}^4_{i,R}}{N_c} \nonumber
\end{eqnarray}
As before, for the beta function in the main text in Eq. (\ref{YkwaRGENcfntegen2}), we rescaled the couplings as $\{\dfrac{g^2_{i}}{4\pi^2}, \dfrac{\lambda}{4\pi^2}  \}\rightarrow \{g^2_i,\lambda \}$ and used the rescaled $g_{i}$ and $\lambda$ for the renormalized couplings. Also, the RG flow is studied in terms of the flow parameter $l$, where $l = -\ln \mu$.

\section{Generalized entanglement entropy of Yukawa operators \label{appndxEentrpy}}

In Sec. \ref{sec:entgnlmnt}, we omit the entanglement of the Yukawa operators in the spin-1/2 space while calculating the entanglement entropy. We have argued that the spin entropy is uniform throughout the manifold and so doesn't play a role in differentiating different conformal theories.

In this appendix, we calculate the full entanglement entropy by including spin entropy and show that its contribution is simply an additive constant ($= \ln 2$), independent of both the manifold coordinates and the flavor number.

We define the generalized Yukawa theory in the following form,
\begin{eqnarray}
    \mathcal{O}_{\text{Y}} = \phi^{*}\sum_{i,j=1}^{\mathcal{N}} \sum_{\alpha,\beta \in \{\uparrow,\downarrow\}}  \tilde{A}_{i\alpha,j\beta} \psi_{i\alpha}^{T}\psi_{j\beta} + \text{h.c}.
\end{eqnarray}
In our discussions, we further restrict ourselves to the limit where $\tilde{A}_{i\alpha,j\beta}$ is factorizable. It is given by,
\begin{eqnarray}
   \tilde{A}_{i\alpha,j\beta} = A_{i,j} \otimes i (s_y)_{\alpha, \beta} 
\end{eqnarray}
where $A_{ij}$, defined in Eq. (\ref{Ykwaoprtr}) encodes the entanglement data of the operator in the flavor subspace.

The full reduced density operator takes the form,
\begin{eqnarray}
    \rho &=& \frac{1}{\text{Tr}[\tilde{A}\tilde{A}^{\dagger}]} \tilde{A}\tilde{A}^{\dagger} \nonumber \\
    &=& \frac{1}{\text{Tr}[AA^{\dagger}]}
    A A^{\dagger} \otimes \rho_{s} \nonumber \\
    &=& \rho_F \otimes \rho_s
\end{eqnarray}
Here $\rho_s$ is the reduced density matrix in the spin space. 

For a Yukawa operator in Eq. (\ref{Ykwaoprtr}), $\rho_s$ takes the desired form of
\begin{eqnarray}
    \rho_s = \frac{1}{2} \left(\begin{array}{cc}
       1  & 0 \\
       0  & 1
    \end{array} \right)
\end{eqnarray}
The generalized entanglement entropy becomes,
\begin{eqnarray}
    S &=& - \text{Tr}\left[ \rho \, \text{ln} \, \rho \right] \nonumber \\ 
    &=& S_F + \ln 2,
\end{eqnarray}
where $S_F$ is the flavor space entanglement entropy, defined in Eq. (\ref{Eentrpy}). Hence, we showed that the effect of spin entropy is simply an additive background constant to the generalized entanglement entropy.

\bibliography{ykwabib}

\end{document}